\documentclass[useAMS,usenatbib,aas_macros]{mn2e}

\usepackage{amssymb,amsmath}
\usepackage{graphicx}
\usepackage{color}
\usepackage{flafter}
\usepackage{placeins}
\usepackage{subfigure}
\usepackage{booktabs}
\usepackage{array}
\usepackage{natbib}
\usepackage{aas_macros}
\usepackage{multirow}
\usepackage[T1]{fontenc} 
\usepackage{aecompl}
\usepackage{csvsimple}
\usepackage{adjustbox}
\usepackage{caption}
\title[Breaking the disc-halo degeneracy]{Constraining the dark matter content of NGC~1291 using hydrodynamic gas response simulations}
\author[Fragkoudi et al.]{F. Fragkoudi$^{1,2}$\thanks{E-mail:
francesca.fragkoudi@obspm.fr}, E. Athanassoula$^{2}$, A. Bosma$^{2}$  \\
$^{1}$GEPI, Observatoire de Paris, CNRS, Universit\'{e} Paris Diderot, 5 Place Jules Janssen, 92195 Meudon, France \\
$^{2}$Aix Marseille Univ, CNRS, LAM, Laboratoire d'Astrophysique de Marseille, Marseille, France\\
}
\begin{document}

\date{}

\pagerange{\pageref{firstpage}--\pageref{lastpage}} \pubyear{2014}

\maketitle

\label{firstpage}

\begin{abstract}

We present a pilot study on the nearby massive barred galaxy NGC~1291, in which we use dynamical modelling to constrain the disc mass-to-light ratio (M/L), thus breaking the degeneracy between the baryonic and dark matter in its central regions. We use the gas, specifically the morphology of the dust lanes on the leading side of the bar, as a tracer of the underlying gravitational potential. We run a large number of hydrodynamic gas response simulations, in potentials obtained directly from near-infrared images of the galaxy, which have three free parameters:  the M/L, the bar pattern speed and the height function. We explore the three-dimensional parameter space, by comparing the morphology of the shocks created in the gas response simulations with those of the observed dust lanes, and find the best-fitting models;
these suggest that the M/L of NGC~1291 agrees with that predicted by stellar population synthesis models in the near-infrared ($\approx$0.6\,$M_{\odot}/L_{\odot}$), which leads to a borderline maximum disc for this galaxy. Furthermore, we find that the bar rotates fast, with a corotation radius which is $\leq$ 1.4 times the bar length.
\end{abstract}



\section{Introduction}
Since the discovery in the 70's that rotation curves derived from HI observations -- which extend well beyond the optical disc of galaxies -- remain flat \citep{Roberts1976,Bosma1978,Bosma1981}, the nature and distribution of non-baryonic dark matter in galaxies has been a source of debate in the scientific community. One of the main issues with determining the non-baryonic distribution of matter in the central regions of galaxies is what has become known as the ``disc-halo degeneracy''. This degeneracy arises due to the fact that rotation curve decompositions depend critically on the mass-to-light ratio (M/L or $\Upsilon$) of the stellar disc. \cite{vanAlbadaetal1985} showed that the rotation curve of a galaxy can be fit equally well with a variety of dark matter halo and stellar disc components, from models with barely any disc contribution, to models where the disc contributes as much as possible to the rotation curve -- without creating a hole in the dark matter halo; this is also known as the ``maximum disc hypothesis''. \cite{Sackett1997} gave a more concrete definition of the maximum disc by stating that a disc is maximum if it contributes 85\%$\pm$10\% of the rotation curve at $r$=2.2$h_r$, where $h_r$ is the scalelength of the disc. However, there is no conclusive evidence to either prove or disprove the maximum disc hypothesis and in fact there are a number of arguments both for \citep{ABP1987,Sackett1997,Weiner2001} and against it \citep{KuijkenGilmore1991,CourteauRix1999}, for the Milky-Way \citep{BissantzGerhard2002,Bissantzetal2003,Hamadacheetal2006,Tisserandetal2007} as well as for external galaxies \citep{Bottema1993,Trottetal2002,Gnedinetal2007,Kranzetal2003}. The lack of knowledge regarding the dark matter content in the central regions of galaxies is of course a major roadblock for testing theories of galaxy formation and evolution.

One way to obtain an estimate of the disc M/L ratio, and therefore break the disc-halo degeneracy, is from stellar population synthesis (SPS) models. These however require a number of assumptions which depend on the detailed star formation history (SFH), the modelling of late phases of stellar evolution and the initial mass function (IMF), quantities which in general contain a large amount of uncertainty, and which propagate through to the determination of the physical parameters of galaxies \citep{BelldeJong2001,Conroyetal2009}.

Another way to break the degeneracy between the disc and the dark matter halo is by using dynamical estimates of the M/L ratio of the disc. In this study we employ a method which uses the gas in galaxies as a tracer of the underlying gravitational potential, and which we refer to as the hydrodynamic gas response method. This method has already been used in studies in the past for both barred galaxies \citep{LLA1996,Weiner2001,Weiner2004, Perezetal2004, Zanmar-Sanchezetal2008} and spiral galaxies \citep{Kranz2001,Kranzetal2003}. This type of modelling has also been applied in order to determine the pattern speed of barred galaxies \citep{Perezetal2004,Sanchez-Menguianoetal2015}. 
The method relies on the simple principle that gas will follow circular orbits when placed in an axisymmetric potential, while, when strong non-axisymmetries are present, it is forced to shock, thus creating the dust lanes observed in many barred galaxies \citep{Prendergast1983, Athanassoula1992b}. The strength and morphology of these shocks depends, among other things, on the ratio between the amount of baryonic to dark matter. Therefore, by tuning the M/L of the disc, such that the models reproduce the morphology and kinematics of the gas in the observations, we can also constrain the amount of dark matter.

In addition to its dependence on the M/L of the disc, the morphology and kinematics of the gas also depend on the bar pattern speed and on the height function of the disc. The effect of the latter has not been extensively studied in the past, however we show in the following sections that it plays a crucial role on the strength and shape of the shocks created in the gas. Due to the dependence of the gas morphology on these three parameters, we need to run a large number of simulations, in order to explore the full three dimensional parameter space, to find the model which best reproduces the observations.

The paper is structured as follows: in Section \ref{sec:obs1291} we describe the observational data we have for NGC~1291, which we use to construct the models. In Section \ref{sec:constructing_model} we present the dynamical models and the gas response simulations. In Section \ref{sec:res} we describe the main results of the paper, the trends that arise from changing the three free parameters and the best fit models. In Section \ref{sec:discussion4} we discuss the findings of this study in the context of previous results in the literature and in Section \ref{sec:summary4} we give a summary of the main results.

\section{Observations of NGC1291}
\label{sec:obs1291} 
\begin{figure}
\centering
\includegraphics[width=0.48\textwidth]{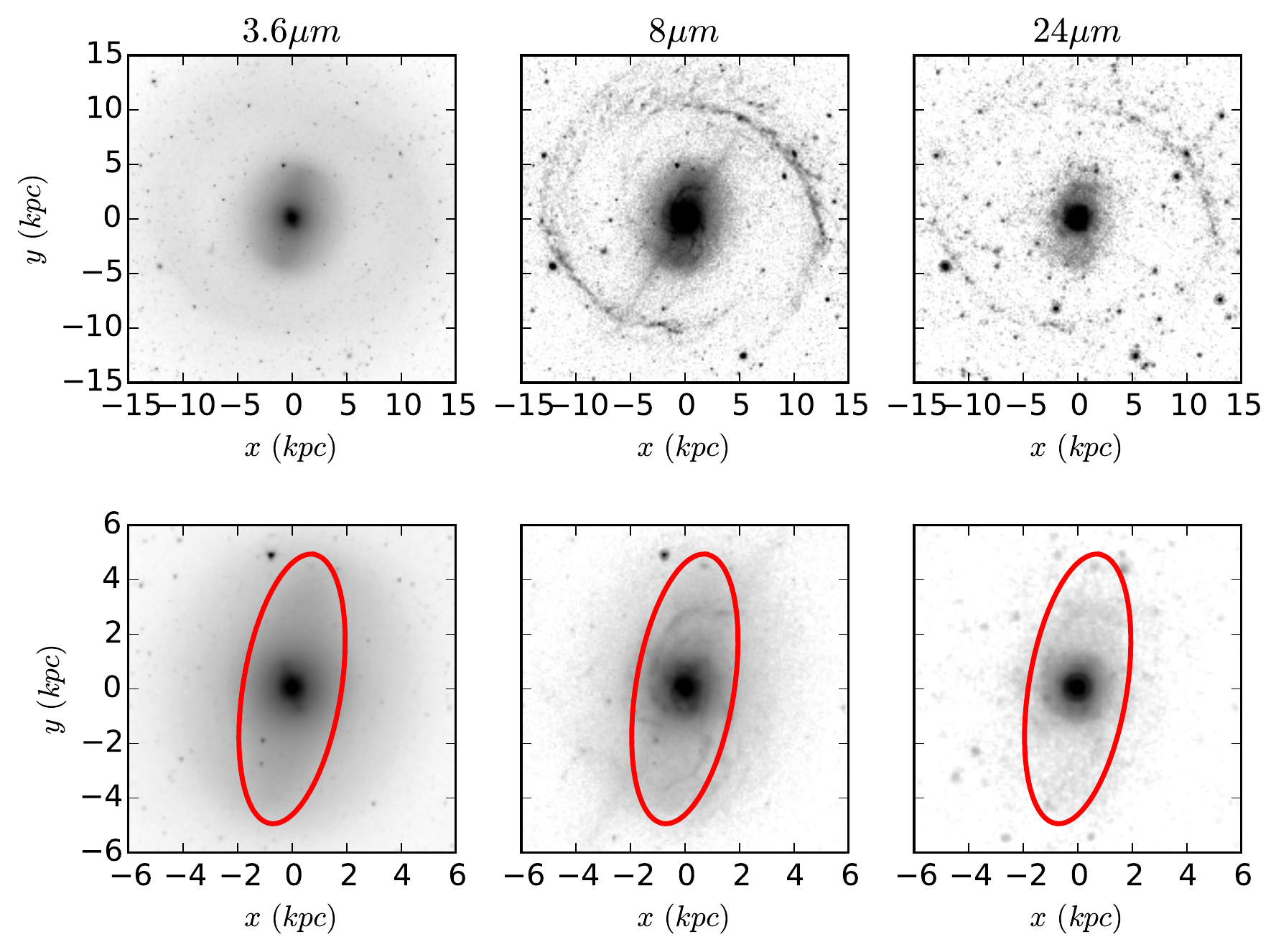}
\caption{From left to right: the 3.6, 8 and 24$\mu$m images for NGC1291. In the top panels the whole galaxy is shown, including the outer ring, while in the bottom panels we zoom in on a region encompassing the bar. The red line outlines the bar as obtained from two-dimensional decompositions of the 3.6$\mu m$ image.} 
\label{fig:ngc1291_36_8}
\end{figure}

\begin{table}
\centering
\caption{Some basic properties of NGC~1291}
\label{tab:info}
\begin{tabular}{ c | c } 
Property & Value \\ \hline \hline
Classification & SB0/a \\
Inclination (deg) & 12 \\
Distance & 9.7\,Mpc \\
1'' & 47\,pc \\
disc scalelength & 5.8\,kpc \\
radius of primary bar & 5.1\,kpc \\
\end{tabular}
\end{table}

NGC~1291 is an early type barred galaxy (SB0/a) \citep{deVaucouleursetal1991} which in addition to the primary bar, contains a smaller nuclear bar, as well as an outer ring, in which there is active star formation \citep{Thilkeretal2007}. The inclination of the galaxy, is at most 12 degrees so it is practically face-on \citep{Bosmaetal2010} and it is at a distance of 9.7\,Mpc \citep{Tullyetal2009}, while the galaxy doesn't seem to be undergoing any interactions. 

The galaxy contains hot gas primarily in the bulge region, as evidenced by X-ray studies \citep{Bregmanetal1995,Hoggetal2001} as well as HI gas, which has a hole in the inner parts of the galaxy where the X-ray gas is found \citep{Bosmaetal2010}.
The fraction of HI gas mass compared to the stellar mass\footnote{Where the stellar mass is obtained using the M/L from S4G \citep{Meidtetal2014,Roecketal2015}, which as we will show further down is a reasonable assumption.} is 2.6\% \citep{Bosmaetal2010} and the mass of the hot gas found in the galaxy is even less \citep{Bregmanetal1995}.

For the work presented here we made use of the 3.6$\mu$m, 4.5$\mu$m,  8$\mu$m and 24$\mu$m images (see Fig. \ref{fig:ngc1291_36_8}) of NGC~1291. 
The 3.6 and 4.5\,$\mu$m images are taken from the Spitzer Survey of Stellar Structure in Galaxies (S$^4$G, \citealt{Sheth2010_s4g}) and are used to obtain the gravitational potential of the galaxy, since these wavelengths are well-suited for tracing the old stellar population, where most of the mass of the galaxy resides. The 3.6$\mu$m image was also used to perform photometric decompositions -- using the BUDDA code \citep{Gadotti2008} -- to obtain the scalelength of the disc and the bar length, which are $h_r$=5.8\,kpc and  $r_B$=5.1\,kpc respectively \citep{Bosmaetal2010}.

To extract information about the dust lanes in the galaxy (see middle and right panels of Figure \ref{fig:ngc1291_36_8}), we use the 8 and 24$\mu$m images from the SINGS survey \citep{Kennicuttetal2003}, which trace the emission due to Polyaromatic Hydrocarbons (PAHs) and hot dust respectively and correlate with the presence of molecular gas \citep{Reganetal2006, Bendoetal2007,Schinnereretal2013}. PAH emission is correlated to the emission from cold dust $\leq$20K \citep{Bendoetal2010} and is likely to occur in shocked regions of the ISM \citep{Tielens2008}. 

Some basic properties of the galaxy are summarised in Table \ref{tab:info}.

\section{Dynamical models \& simulations}
\label{sec:constructing_model}

To obtain a dynamical model of NGC~1291, we need to account for the mass distributions of the different components which make up the potential, such as the stellar (Section \ref{sec:pot}) and dark matter (Section \ref{sec:dmhalo}) component. 
In general one also needs to account for the gaseous component in the galaxy, however, as mentioned in Section \ref{sec:obs1291} , the HI and X-ray gas are negligible in the bar region of NGC~1291 and we therefore use the gas only as a tracer of the underlying potential.

\subsection{Potential of the stellar component}
\label{sec:pot}
The method for calculating the potential due to the stellar component, involves a straightforward three-dimensional integration over the density distribution, which was described in more detail along with tests on the code in \cite{Fragkoudietal2015}. Once a mass map for the galaxy is constructed from the NIR image of the galaxy -- by assigning a M/L -- we then assign a height function and integrate over the density distribution to obtain the gravitational potential. 

\subsubsection{The Mass to Light Ratio}
\label{sec:m/l}

To construct a mass map of the galaxy we first remove contaminating sources, such as foreground stars, apply Fourier smoothing to the image, and then assign a M/L to convert the luminosity to mass.
The M/L in the 3.6$\mu m$ band can be assumed to be 0.6 \citep{Meidtetal2014,Roecketal2015}, \emph{once the contamination by dust has been removed}. To correct for dust contamination, we combine information from the 3.6 and 4.5$\mu$m images, using a formula by \cite{Eskewetal2012} with a new calibration by \cite{Querejetaetal2015}, obtained using a large number of S$^4$G galaxies. The equation for transforming from the 3.6 and 4.5$\mu m$ flux densities ($F_{3.6}$ and $F_{4.5}$ in Jy) to mass (in units of $M_{\odot}$) as given by \cite{Querejetaetal2015} is,

\begin{equation}
\frac{M_\star}{\mathrm{M_\odot}} =  10^{8.35} \times \Big(\frac{F_{3.6}}{\mathrm{Jy}}\Big)^{1.85} \times \Big(\frac{F_{4.5}}{\mathrm{Jy}}\Big)^{-0.85}  \times \Big(\frac{D}{\mathrm{Mpc}}\Big)^2,
\label{eq:ml}
\end{equation}
 
\noindent where the distance $D$ is given in Mpc.

\subsubsection{The Height Function}
\label{sec:hf}

\begin{figure}
\centering
\includegraphics[width=0.49\textwidth]{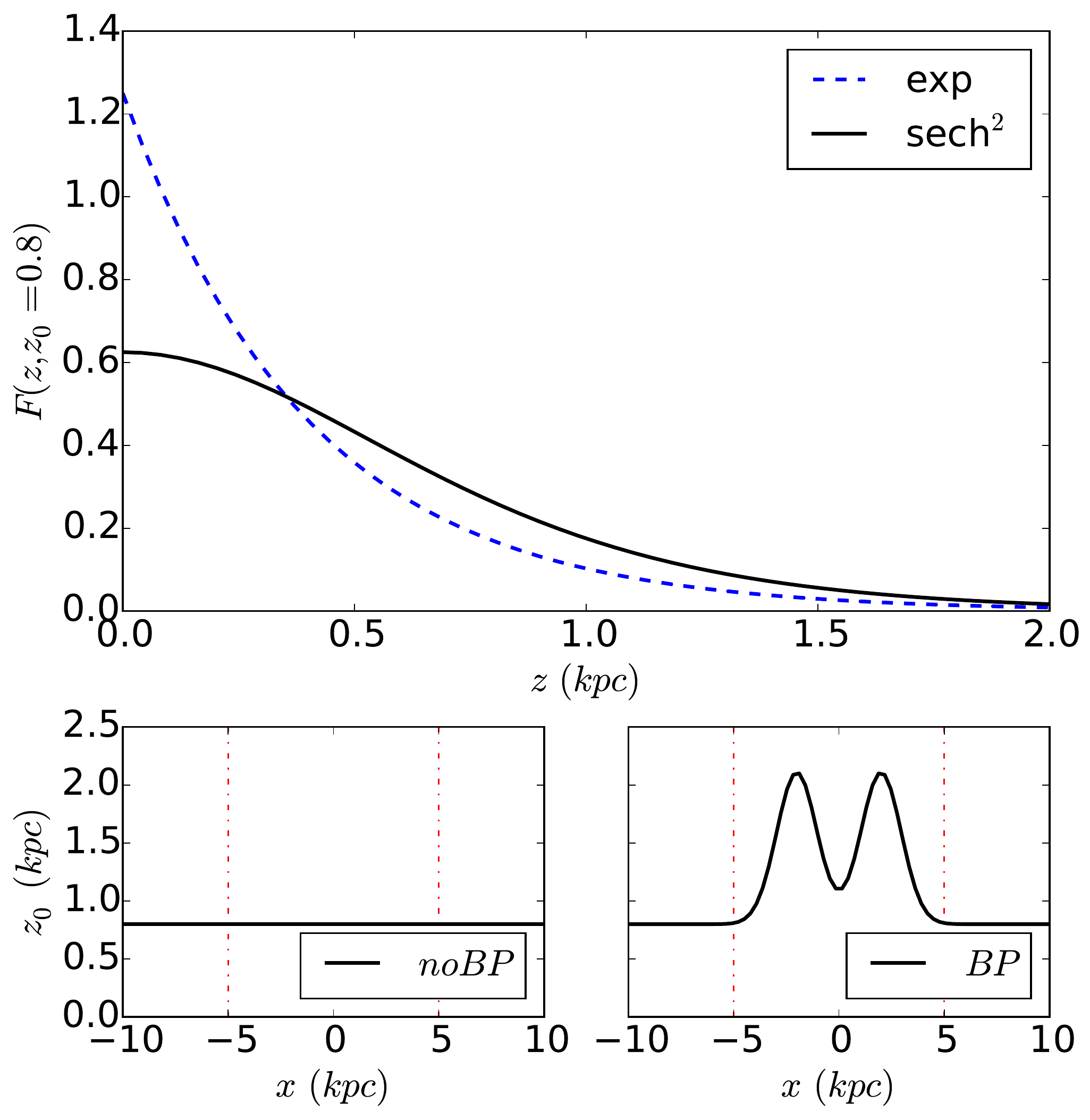} 
\caption{\emph{Top panel:} Isothermal and exponential height functions with equivalent scaleheights ($z_0$ = 2$h_z$=0.8\,kpc). We see that the exponential height function is the more peaked, which leads to higher mass concentrations close to the plane of the galaxy. 
\emph{Bottom panels:} On the left we show the behaviour of $z_0$, the isothermal scaleheight for a disc with constant scaleheight, and on the right we show how $z_0$ varies for a disc with our b/p height function.}
\label{fig:scaleheights}
\end{figure}

We assume a height function, which defines how the density drops off as a function of \emph{z} from the equatorial plane.
The height function and the scaleheight ($z_0$) smooth out the potential, and we therefore need to use the height function which best approximates that of the galaxy we are trying to model. 
Different functional forms can be used for the height function, and in this study, we explore two of the most commonly used ones in the literature, the isothermal sheet model and an exponential height function (see top panel Figure \ref{fig:scaleheights}).

The isothermal-sheet model \citep{vanderkruit1981} is given by: 

\begin{equation}
F(z)=\frac{1}{2z_0}\mathrm{sech}^2\left(\frac{z}{z_0}\right)  ,
\label{eq:sech2}
\end{equation}

\noindent where 1/(2$z_0$) is the normalisation factor and $z_0$ is the scaleheight (also referred to as the isothermal scaleheight when a clearer distinction needs to be made from the exponential scaleheight). 

Studies of edge-on galaxies have shown that the vertical profiles of galaxies can be more peaked than that expected for an isothermal height function and better matched by an exponential distribution \citep{deGrijsetal1997,deGrijs1998}. To check whether our assumptions on the form of the height function would have a significant effect on the results, we ran some additional suites of models with an exponential height function, which is given by,
\begin{equation}
F(z)=\frac{1}{2h_z}\mathrm{exp}\left(-\frac{z}{h_z}\right)  ,
\label{eq:sech2}
\end{equation}
where $h_z$ is the exponential scaleheight which is related to the isothermal scaleheight by $z_0$ = 2\,$h_z$.

We also explore height functions which include the shape of the boxy/peanut (b/p) bulge, since, as was shown in \cite{Fragkoudietal2015}, when a b/p bulge is present in a galaxy it should be modelled, because it can have a significant effect on the model. We use a b/p height function which is a non-separable function of position, given by:
\begin{equation}
F(x,y,z)=\frac{1}{2 z_0(x,y)}\mathrm{sech}^2\left(\frac{z}{z_0(x,y)}\right),
\label{eq:peanut1}
\end{equation}

\noindent where the scaleheight \emph{z}$_0$(x,y) varies like the sum of two two-dimensional gaussians (see bottom right panel of Figure \ref{fig:scaleheights}):

\begin{equation}
\begin{split}
z_0(x,y)= & A_{sim} \exp\left(-\left(\frac{(x-x_0)^2}{2\sigma^2} + \frac{(y-y_0)^2}{2\sigma^2}\right)\right) + \\
& A_{sim} \exp\left(-\left(\frac{(x-x_1)^2}{2\sigma^2} + \frac{(y-y_1)^2}{2\sigma^2}\right)\right) + z_0^{disc} ,
\end{split}
\label{eq:peanut2}
\end{equation}

\noindent where $A_{sim}$ is the maximum scaleheight of the peanut above the disc scaleheight ($z_0^{disc}$) and $\sigma^2$ is the variance of the gaussians. Parameters ($x_0$, $y_0$, $x_1$, $y_1$) give the positions of the maxima of the gaussians. 

To obtain an initial set of values for the parameters of the b/p height function, we fitted the vertical mass distribution of an $N$-body simulation of an isolated galaxy, taken from \cite{Athanassoulaetal2014} (for more details on the fitting, see \cite{Fragkoudietal2015}). 
In what follows in this study, we adopt a fiducial value for the maximum scaleheight of the peanut $A_{fid}$, which is half that obtained by the fit to the simulated galaxy $A_{sim}$ i.e.,
\begin{equation}
A_{fid}=\frac{A_{sim}}{2}=0.65 \,\mathrm{kpc}
\end{equation}
This is done because the height of the b/p, also known as the strength of the b/p, is known to be correlated with the bar strength \citep{Athanassoula2008,MartinezValpuestaAthanassoula2008} -- although there is a large amount of scatter in this relation. The simulated galaxy used to fit the b/p height function develops a strong bar, stronger than that of NGC~1291. We therefore reduce the strength of the b/p for the model of NGC~1291, since such a strong b/p is not likely to be present in more weakly barred galaxies.

\subsection{Dark matter halo potential}
\label{sec:dmhalo}

\begin{figure}
\centering
\subfigure[$f_d$=1.0]{%
	\includegraphics[width=0.22\textwidth]{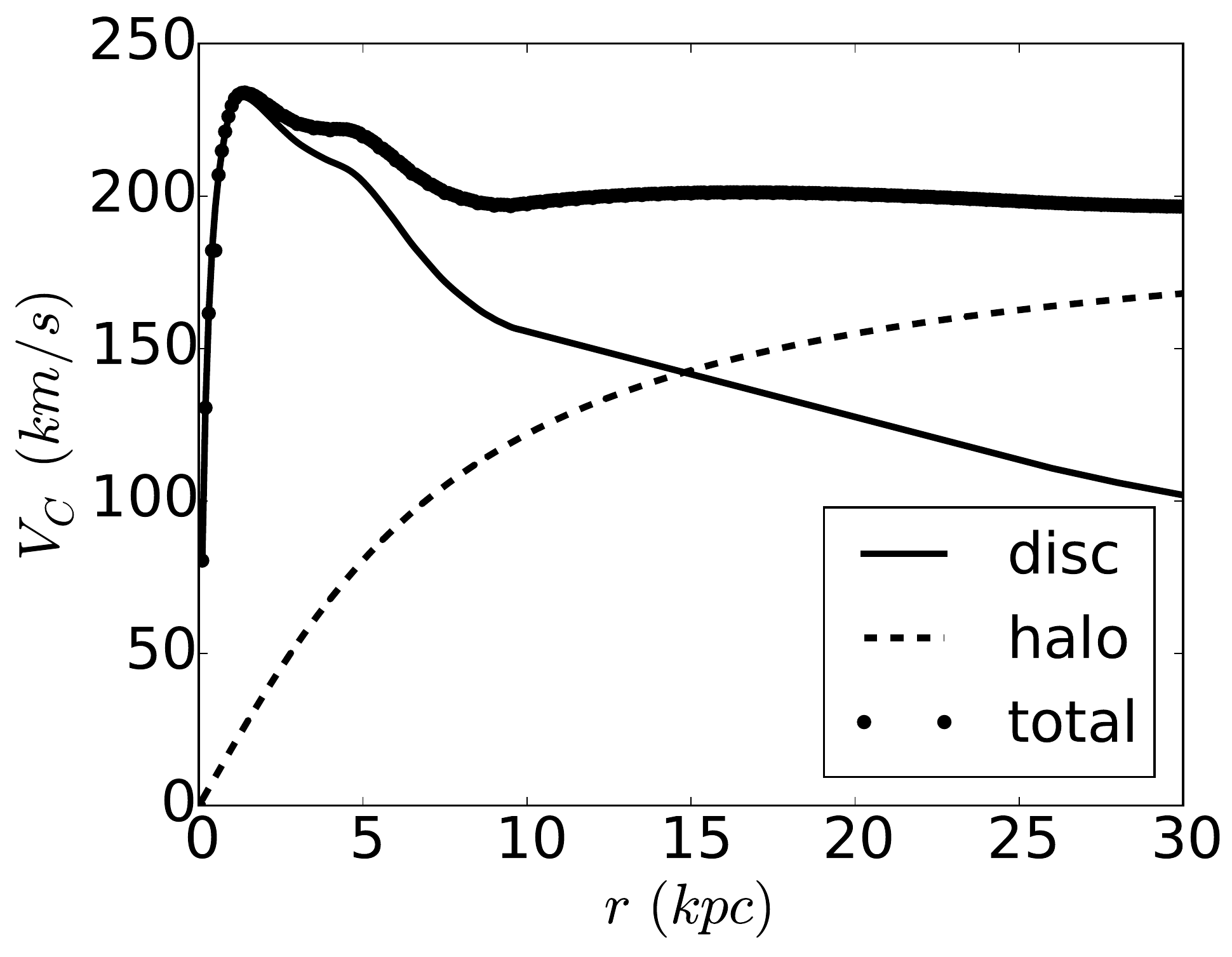}
	\label{fig:mod001}}
\quad
\subfigure[$f_d$=0.75]{%
	\includegraphics[width=0.22\textwidth]{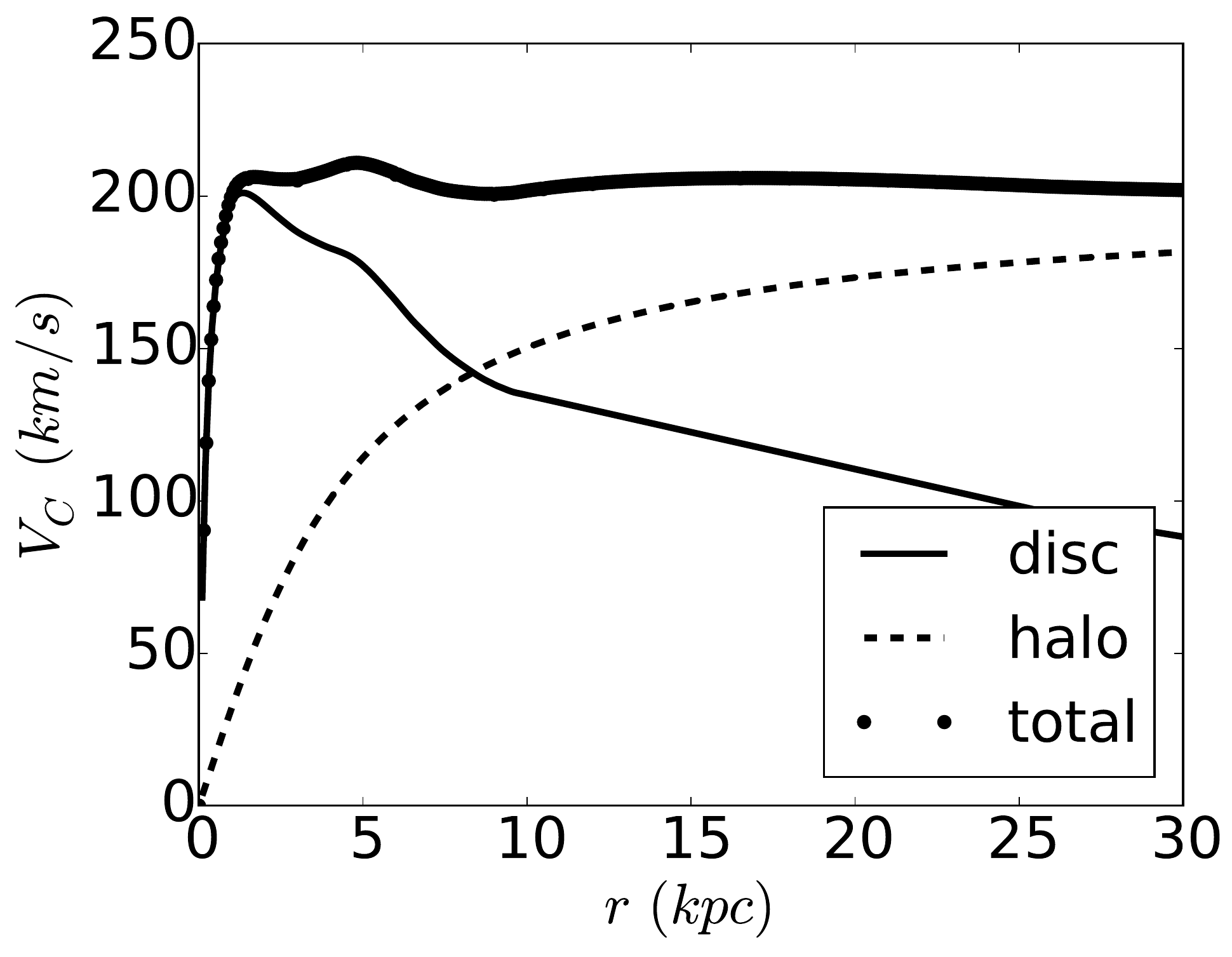}
	\label{fig:mod002}}
\quad
\subfigure[$f_d$=0.5]{%
	\includegraphics[width=0.22\textwidth]{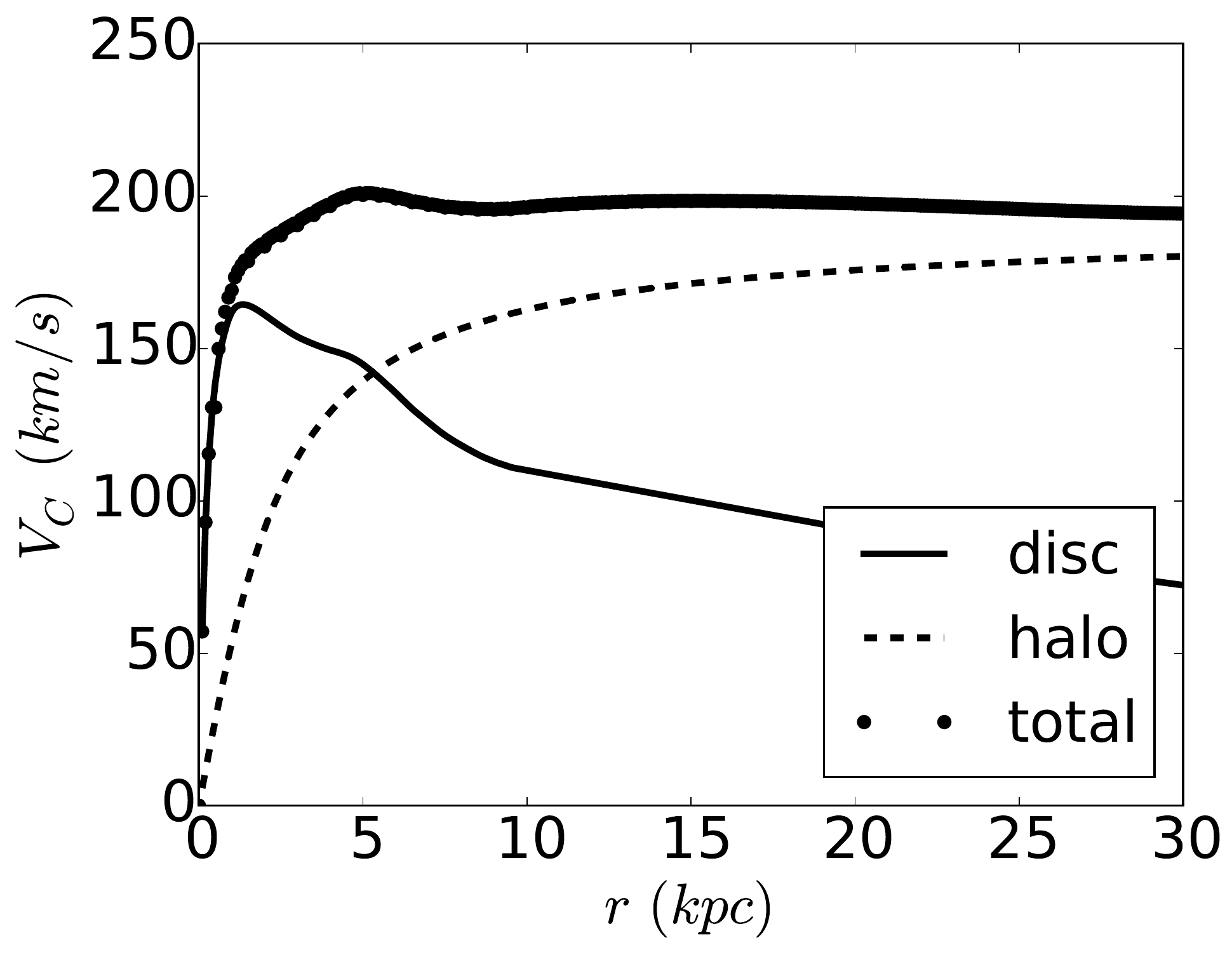}
	\label{fig:mod003}}
\quad
\subfigure[$f_d$=0.25]{%
	\includegraphics[width=0.22\textwidth]{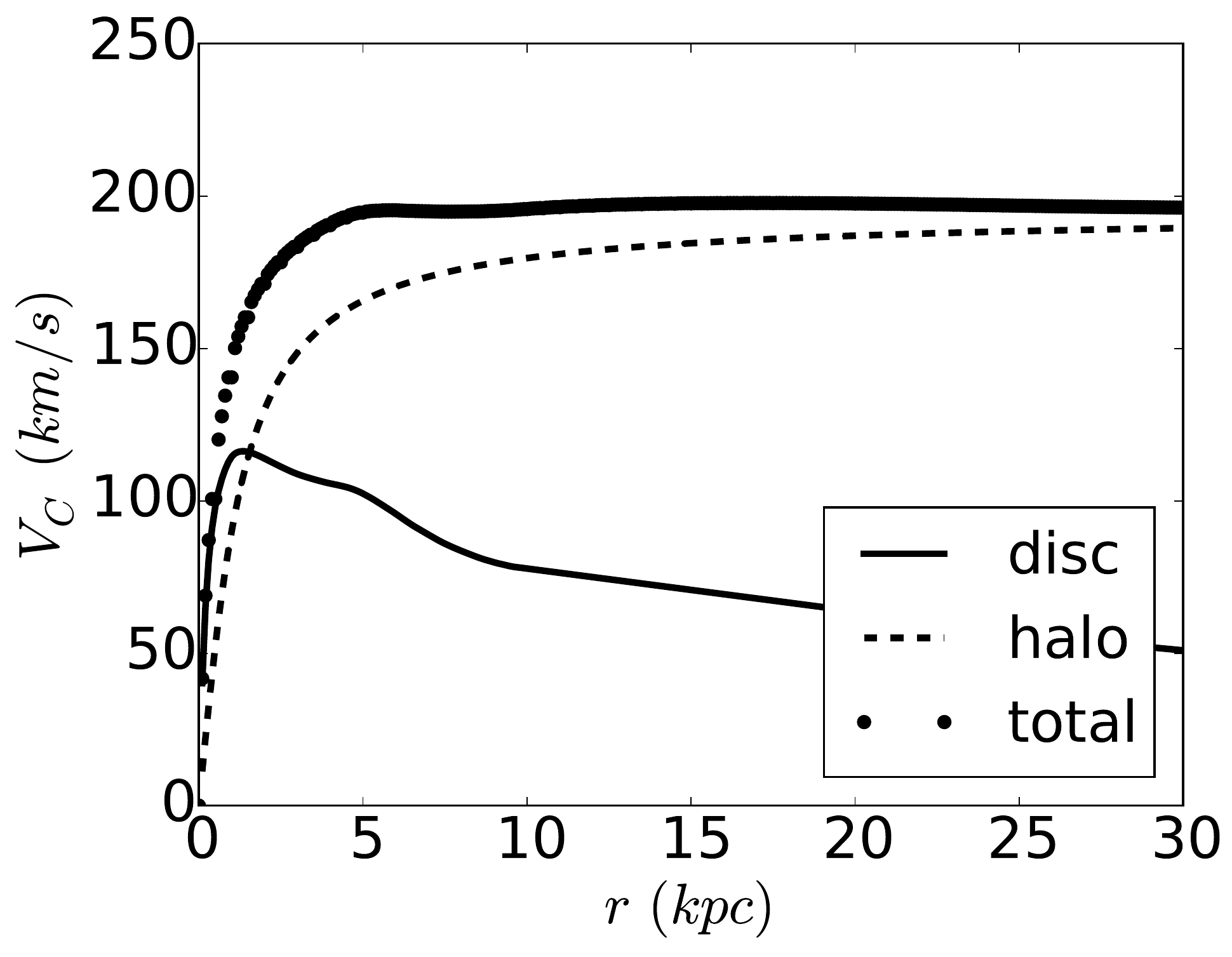}
	\label{fig:mod004}}
\quad
\caption{Rotation curves for models 001 to 004. The M/L is decreasing from 1$\times$$\Upsilon_{3.6}$ to 0.25 $\times$$\Upsilon_{3.6}$ in steps of 0.25. } 
\label{fig:rotcurve}
\end{figure}

The dark matter halo in our models is a spherical pseudo-isothermal halo whose density is given by,

\begin{equation}
\rho(r)=\rho_c \Big[ 1 + \Big(\frac{r}{r_c}\Big)^2\Big]^{-1}
\label{eq:rho_pseudo}
\end{equation}

\noindent where $r_c$, the core radius, and $\rho_c$, the central density, are adjusted such that $V_{flat}$ is the same in all the models. The circular velocity at a given radius $r$, is given by,

\begin{equation}
V(r)=V_{\infty} \Big[ 1 - \frac{r_c}{r}\mathrm{arctan}\Big(\frac{r}{r_c}\Big)\Big]^{1/2}
\label{eq:v_pseudo}
\end{equation}

\noindent where $V_{\infty}$ is the circular velocity at infinity. 
The parameters of the halo are obtained by adjusting the rotation curve of the model, such that it reproduces the rotation velocity in the outer parts ($V_{flat}$), as predicted by the Tully-Fisher (TF) relation for NGC~1291 \citep{McGaughetal2000,McGaugh2012}. 
For NGC~1291 $V_{flat}$ = 200 km/s, and we use this value to construct the rotation curves (see Figure \ref{fig:rotcurve}).

\subsection{The Lagrangian radius}
The Lagrangian radius is the distance from the centre of the galaxy to the Lagrangian points $L_1$ (or $L_2$) which are the locations at which the forces in the rotating frame of reference are exactly zero, i.e. in the rotating frame of reference, the stars in the disc rotate at the same speed as the bar.
The Lagrangian radius is often parametrised by,

\begin{equation}
r_L = \mathcal{R}\, r_B.
\end{equation}

\noindent where $r_L$ is the Lagrangian radius, $r_B$ is the bar semi-major axis and $\mathcal{R}$ is a parameter which is often used to separate bars into fast $\mathcal{R}$ $\leq$ 1.4 and slow $\mathcal{R}$ $\textgreater$ 1.4 bars (e.g. \citealt{DebattistaSellwood2000}).

\subsection{Hydrodynamic Simulations}
\label{sec:sims}

The gas dynamic (or gas response) simulations were run using the hydrodynamic Adaptive Mesh Refinement (AMR) grid code RAMSES \citep{Teyssier2002}. The simulations are run in two dimensions, since, to zeroth order, gas is confined to a thin plane in disc galaxies, which allows us to therefore decrease the complexity of the problem.

The initial conditions of the simulations consist of a two-dimensional axisymmetric gaseous disc, with a homogeneous density distribution, in hydrostatic equilibrium. The disc is tapered using an exponential tapering function. The gas is modelled as an isothermal gas with adiabatic index 5/3 which corresponds to HI gas (neutral atomic hydrogen), and we give a value of 10km/s to the sound speed of the gas, which corresponds to characteristic temperatures of the interstellar medium.

In order to avoid transients in the disc, we introduce the non-axisymmetric potential over a finite period of time, which was found empirically as $\sim$3 bar rotations, in accordance with previous studies \citep{Athanassoula1992b,PatsisAthanassoula2000}.

\cite{Sormanietal2015} found that the gas flows in non-axisymmetric potentials are sensitive to the resolution of the grid used in these types of simulations. After carrying out a number of tests on analytic models containing a disc, bar and bulge, taken from \cite{Athanassoula1992a}, we determined that the location of the shocks remains the same (within 5\%) for varying resolutions, while the central region of the simulation is indeed sensitive to resolution. For more details on this we refer the reader to Appending A. For the purposes of this study our simulations have a box size of 60\,kpc, and a maximum resolution in the shock regions of 30\,pc.

\subsection{Varying the free parameters}

The goal of this work is to find the best fit between the gas response simulations and the observations, by exploring the allowed parameter space, by varying the three free parameters of the models, i.e. the disc \emph{M/L}, the \emph{Lagrangian radius} and the \emph{height function} of the disc.

In total we ran approximately 300 models, where we varied the M/L from 1.5 to 0.25 $\Upsilon_{3.6}$ in steps of 0.25, the Lagrangian radius $r_L$ = $\mathcal{R}$ $r_B$ where $\mathcal{R}$ was varied between 1 and 2 in steps of 0.2, and various height functions and scaleheights were tested.

\paragraph*{The Mass-to-Light ratio}
The fiducial value of the stellar M/L ($\Upsilon_{3.6}$) is calculated as explained in Section \ref{sec:m/l}. We then change this value, by assuming different fractions of the fiducial $\Upsilon_{3.6}$ according to,
\begin{equation}
\Upsilon = f_d\Upsilon_{3.6}.
\end{equation}
We then re-calculate the disc rotation curve of the galaxy each time and add a dark matter halo with parameters such that the rotation curve in the outer parts of the galaxy matches the value predicted by the TF relation, as shown in Figure \ref{fig:rotcurve}.


\paragraph*{The Lagrangian radius}

By varying the Lagrangian radius while keeping all the other parameters of a model constant, we effectively vary the pattern speed of the bar. 
There are theoretical arguments which show that self consistent bars cannot have a corotation radius smaller than the bar length \citep{Contopoulos1980,Athanassoula1980}, i.e. $\mathcal{R}$ cannot be smaller than 1. On the other hand, what controls the upper limit of the Lagrangian radius is the angular momentum transfer in the galaxy. Therefore the slow-down of the bar will depend on the amount of resonant material available which is able to absorb angular momentum from the inner parts of the discs \citep{Athanassoula2003}.

Previous studies of the shape of dust lanes in gas response simulations have set limits to $\mathcal{R}$, placing the corotation radius $r_L$= (1.2$\pm$0.2)$r_B$ \citep{Athanassoula1992b}, however there is still no consensus in the literature regarding the value of $\mathcal{R}$, and indeed whether it is the same for both early and late-type barred spirals.

\paragraph*{The height function}
The scaleheight of the disc plays an important role in the strength of the forces in the plane of the galaxy \citep{Fragkoudietal2015}. We vary the height function of the disc by assuming two functional forms for the height functions -- isothermal and exponential. Additionally, we explore the effect of adding a b/p height function on the results, i.e. changing the scaleheight at certain locations in the disc.
For the models with the flat height functions, we vary the scaleheight around this value -- from $z_0$=0.5 to $z_0$=1.5\,kpc (and the equivalent exponential scaleheight).

\section{Results}
\label{sec:res}

\begin{figure}
\centering
\includegraphics[width=0.49\textwidth]{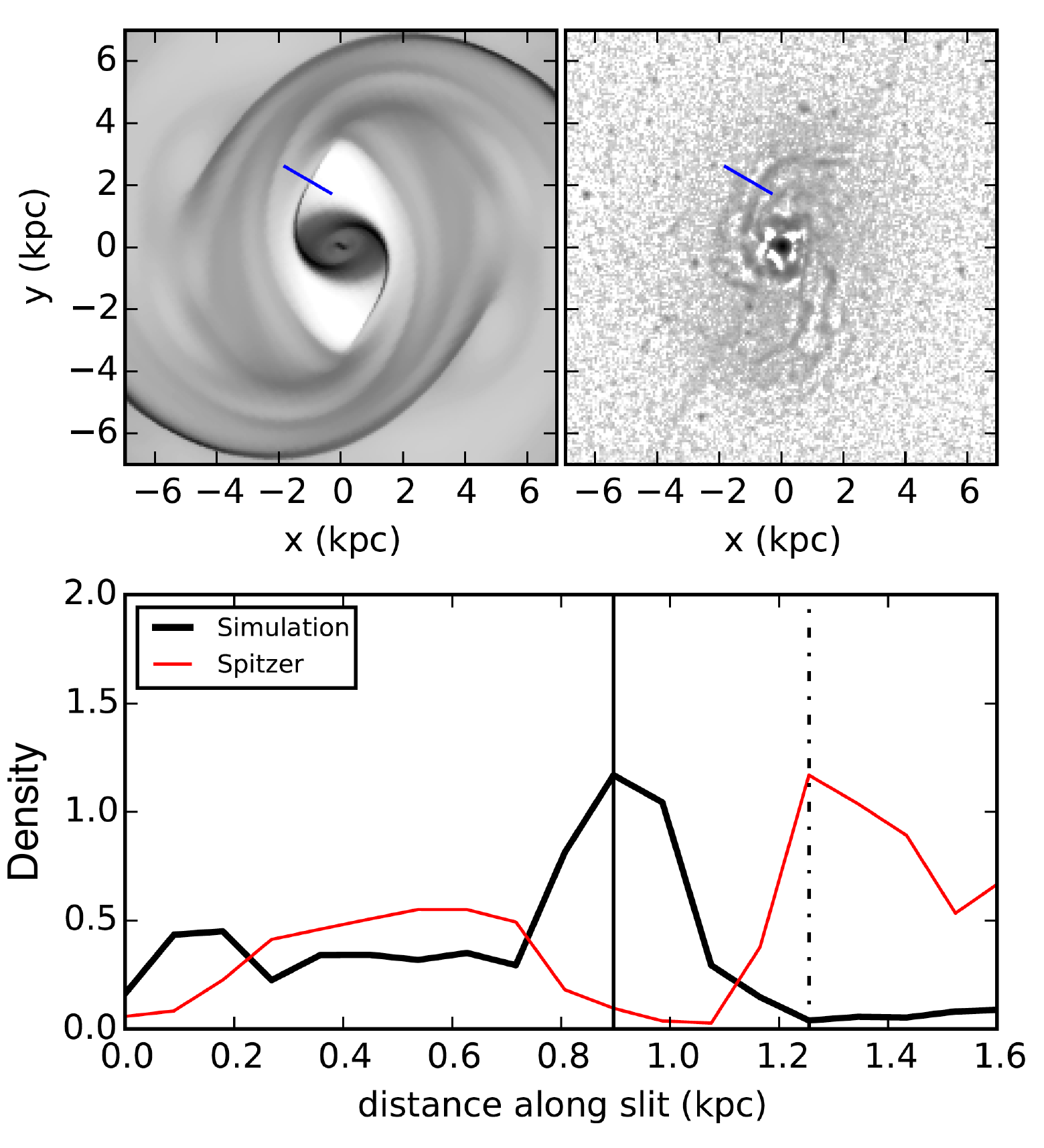} 
\caption{Comparison between the location of the shocks in the models and the observations. A pseudo-slit is placed perpendicular to the shock in the simulation (top left) and the observed unsharp masked 8$\mu$m image (top right). In the bottom panel we show the density along the slit in the two images as a function of distance $l$ along the slit. The shock loci are defined by the maximum density distribution of the simulation and the 8$\mu m$ image, and $\Delta l$ is defined as the distance between the two shock loci.}
\label{fig:mod005_slit}
\end{figure}

To quantitatively compare the models to the data, we place pseudo-slits perpendicular to the shock loci in both the gas flow simulations and the observed image, where the modelled gas flow is rotated to match the rotation of the observed image. As shown in Figure \ref{fig:mod005_slit} we trace the density jumps perpendicular to the slit, where the maxima in the density indicate the location of the shocks. The distance between the location of the shocks in the simulations and observations is denoted by $\Delta l$, which is a proxy to the goodness of fit of the models. 

\subsection{Effect of M/L}
\label{sec:resML}
\begin{figure*}
\centering
\includegraphics[width=1\textwidth]{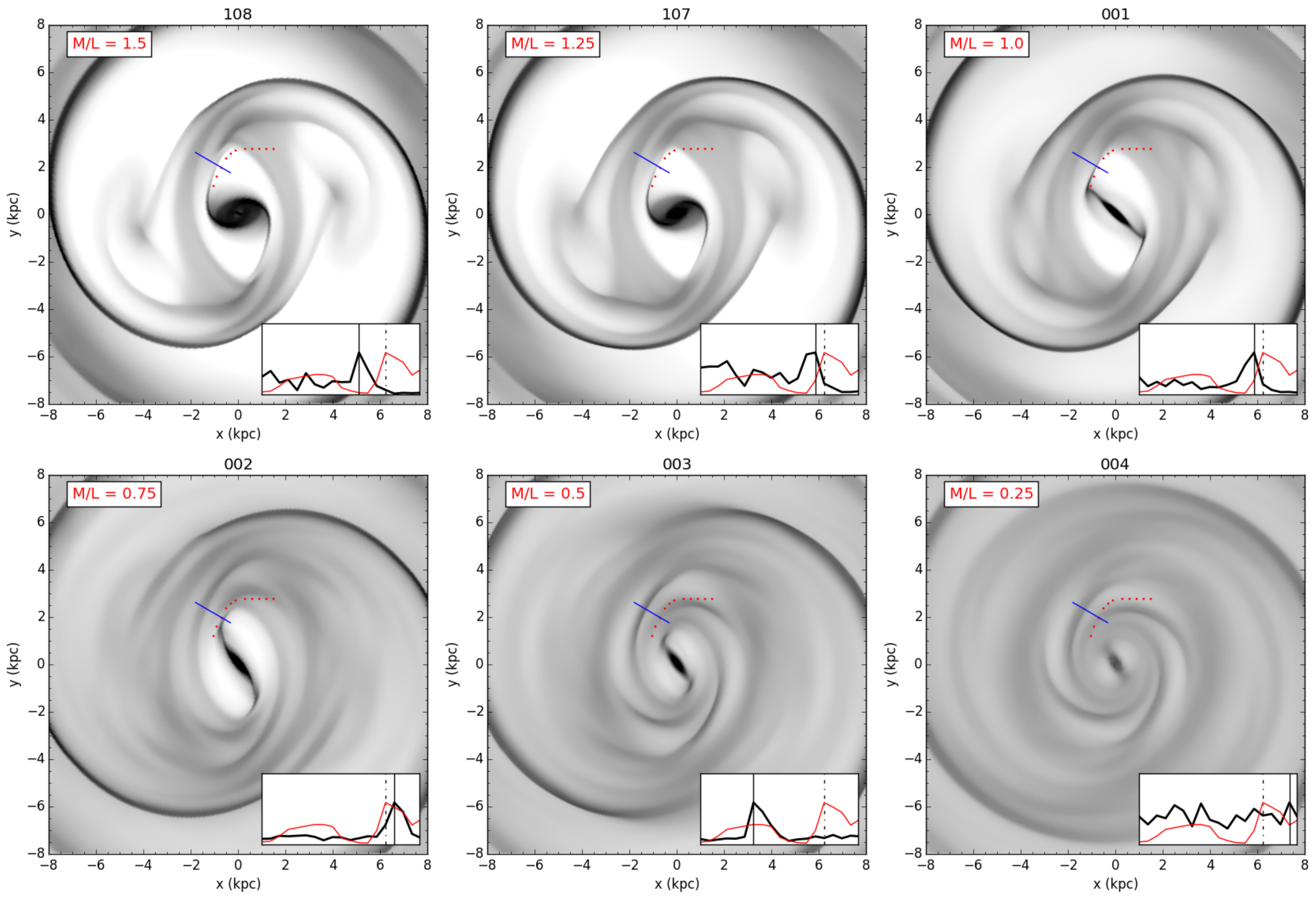} 
\caption{The gas response to models with different M/L: The M/L decreases from top left to bottom right from 1.5 to 0.25$\Upsilon_{3.6}$ (as indicated in the top left corner of the plots). The rest of the parameters are the same for each model, namely the height function is an isothermal disc with scaleheight $z_0$=0.8\,kpc and Lagrangian radius $r_L$=1.2$r_{B}$. The blue line shows the location of the pseudo-slit in the simulation and the red dots show the loci of the shock in the 8$\mu m$ Spitzer image of NGC 1291. The insets show the density along the pseudo-slit for both the simulation (thick black line) and the 8$\mu m$ Spitzer image (thin red line), and are provided to guide the eye on where the strongest shocks occur with respect to the pseudo-slit. The vertical lines in the insets show the shock loci along the slit for the models (solid line) and the observed image (dashed-dotted line).}
\label{fig:masstolight}
\end{figure*}

Decreasing the M/L of the stellar disc has a significant impact on the strength and shape of the shocks formed in the gas, as can be seen in Figure \ref{fig:masstolight}, where we decrease the M/L from 1.5 to 0.25 $\times \Upsilon_{3.6}$ in steps of 0.25. By reducing the M/L and accordingly increasing the contribution of the dark matter halo the shocks become progressively weaker, and the shape of the shocks becomes rounder. Additionally the amount of gas present in the bar region increases since the bar is not strong enough to deplete the gas there. This occurs since we are effectively reducing the non-axisymmetric perturbation in the model, and increase the axisymmetric component. The strength of the non-axisymmetric perturbation determines the strength of the shocks, since by reducing the bar potential we effectively increase the extent of the $x_2$ family of periodic orbits, while the $x_1$ orbits become less cusped; therefore the gas tends to stay on more circular orbits. 

\subsection{Effect of Lagrangian radius}
\label{sec:resPS}

\begin{figure*}
\centering
\includegraphics[width=1\textwidth]{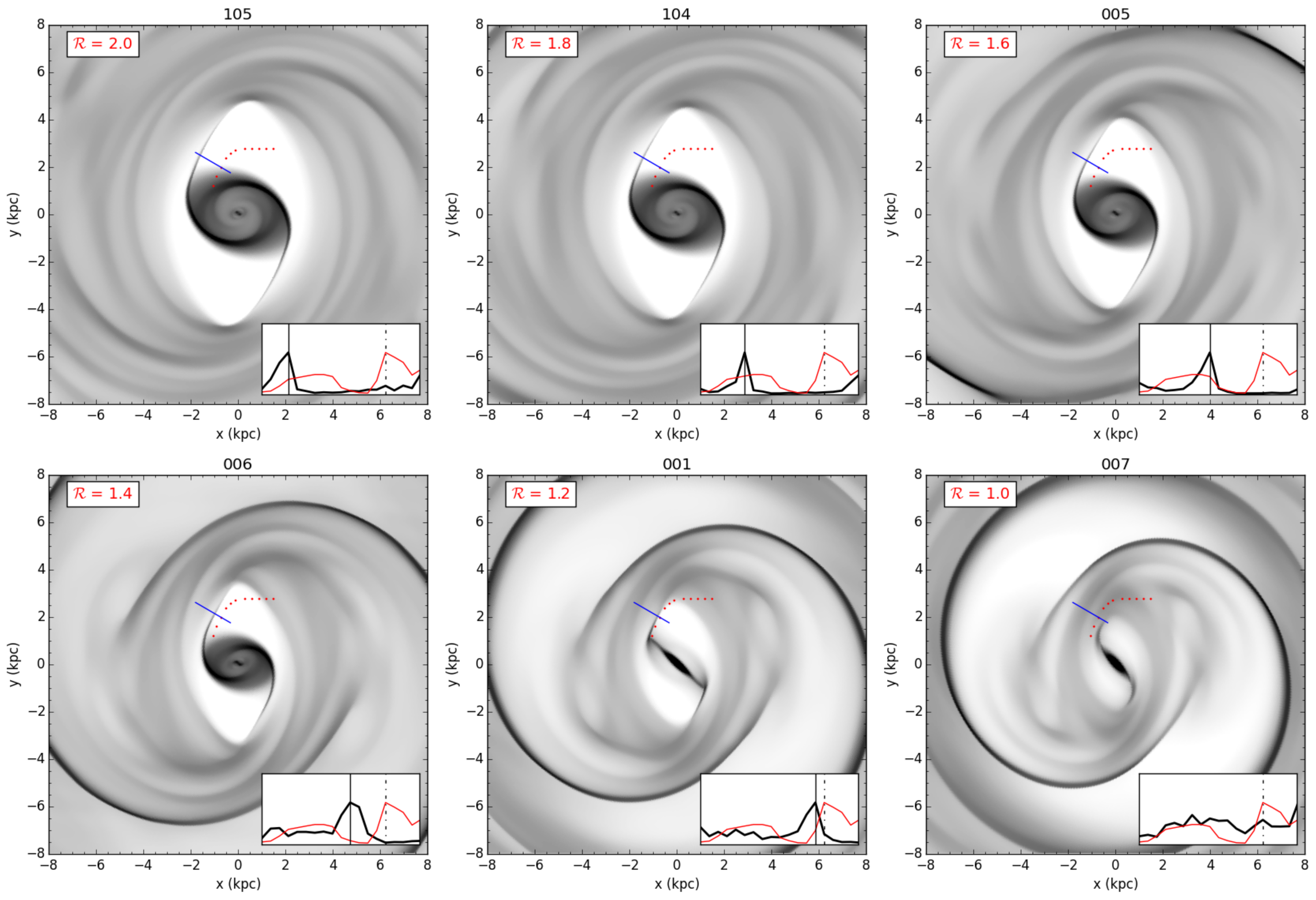}
\caption{As in Figure \ref{fig:masstolight} but for different Lagrangian radii (M/L is equal to $\Upsilon_{3.6}$ in all the models).}
\label{fig:pattern_speed}
\end{figure*}

By changing the pattern speed of the bar, the location of the Lagrangian (or more loosely, the corotation) radius in the disc changes. 
In Figure \ref{fig:pattern_speed}, we show the effect of changing the Lagrangian radius, while leaving all other parameters in the models the same. By increasing the Lagrangian radius, i.e. reducing the pattern speed of the bar, the distance of the shocks from the centre of the bar increases -- in accordance with previous results in the literature \citep{Athanassoula1992b,Sanchez-Menguianoetal2015} -- while the shocks formed at the leading edge of the bar are also stronger.  
For very high pattern speeds the shocks move towards the inside of where the dust lanes occur (see model 007 in Figure \ref{fig:pattern_speed}). The strength of the shocks also decreases for high pattern speeds and therefore the density contrast in the gas also decreases.

We already start to see an issue between changing the Lagrangian radius and the M/L, in that there are similarities in the ways changing the M/L and the pattern speed affect the gas flow. In both cases, i.e. by decreasing the M/L and the Lagrangian radius, the shocks are weakened and their shape becomes rounder, which leads to degeneracies between the different models (for example, compare model 002 of Figure \ref{fig:masstolight} and model 007 of Figure \ref{fig:pattern_speed}).

\subsection{Effect of height function}
\label{sec:resHF}
\begin{figure*}
\centering
\includegraphics[width=1\textwidth]{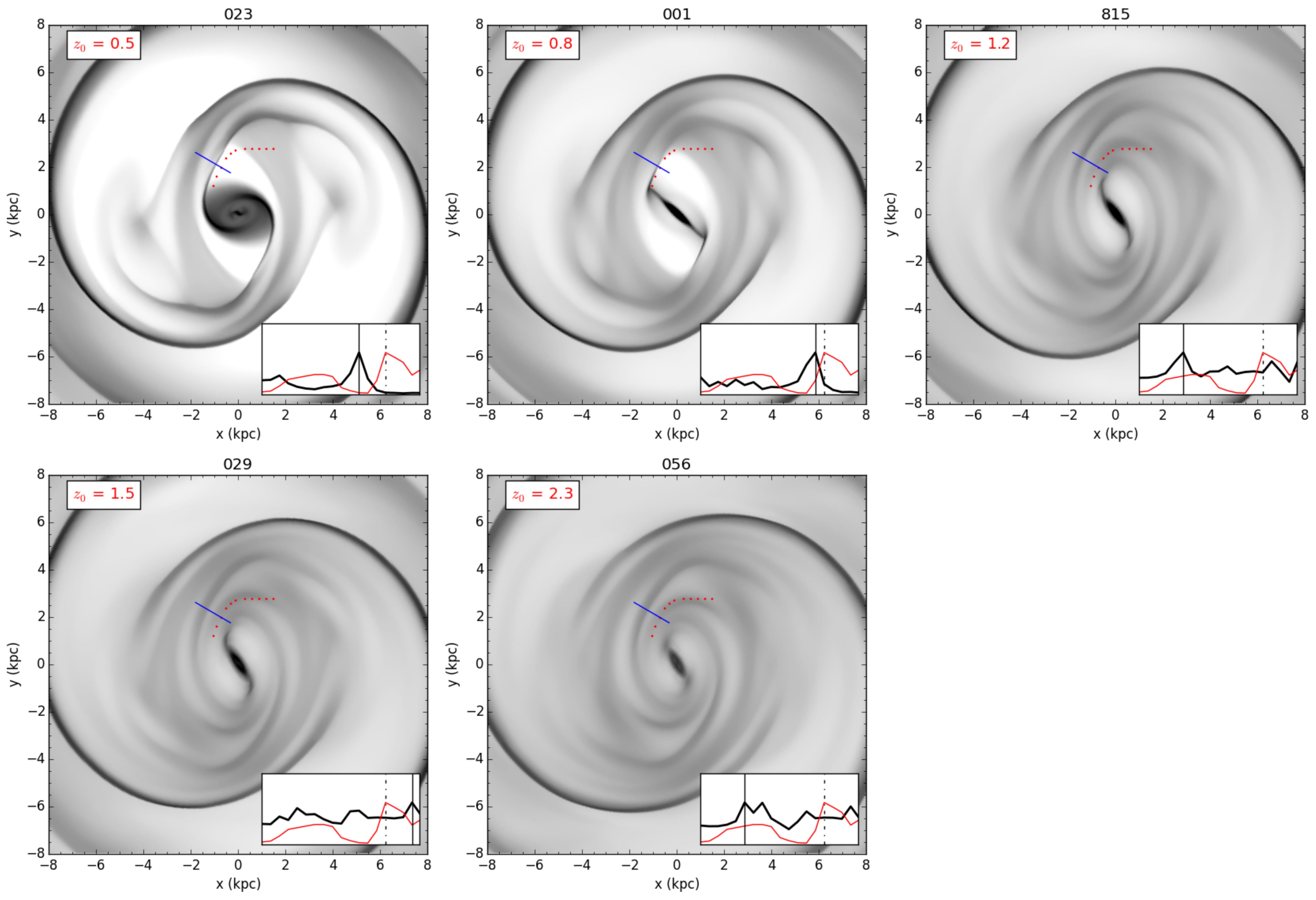} 
\caption{As in Figure \ref{fig:masstolight} but changing the scaleheight of the models. The other parameters of the models are the same, namely M/L = $\Upsilon_{3.6}$ and $\mathcal{R}$ =1.2.}
\label{fig:heightfunction}
\end{figure*}

\begin{figure}
\centering
\subfigure[Isothermal]{%
	\includegraphics[width=0.22\textwidth]{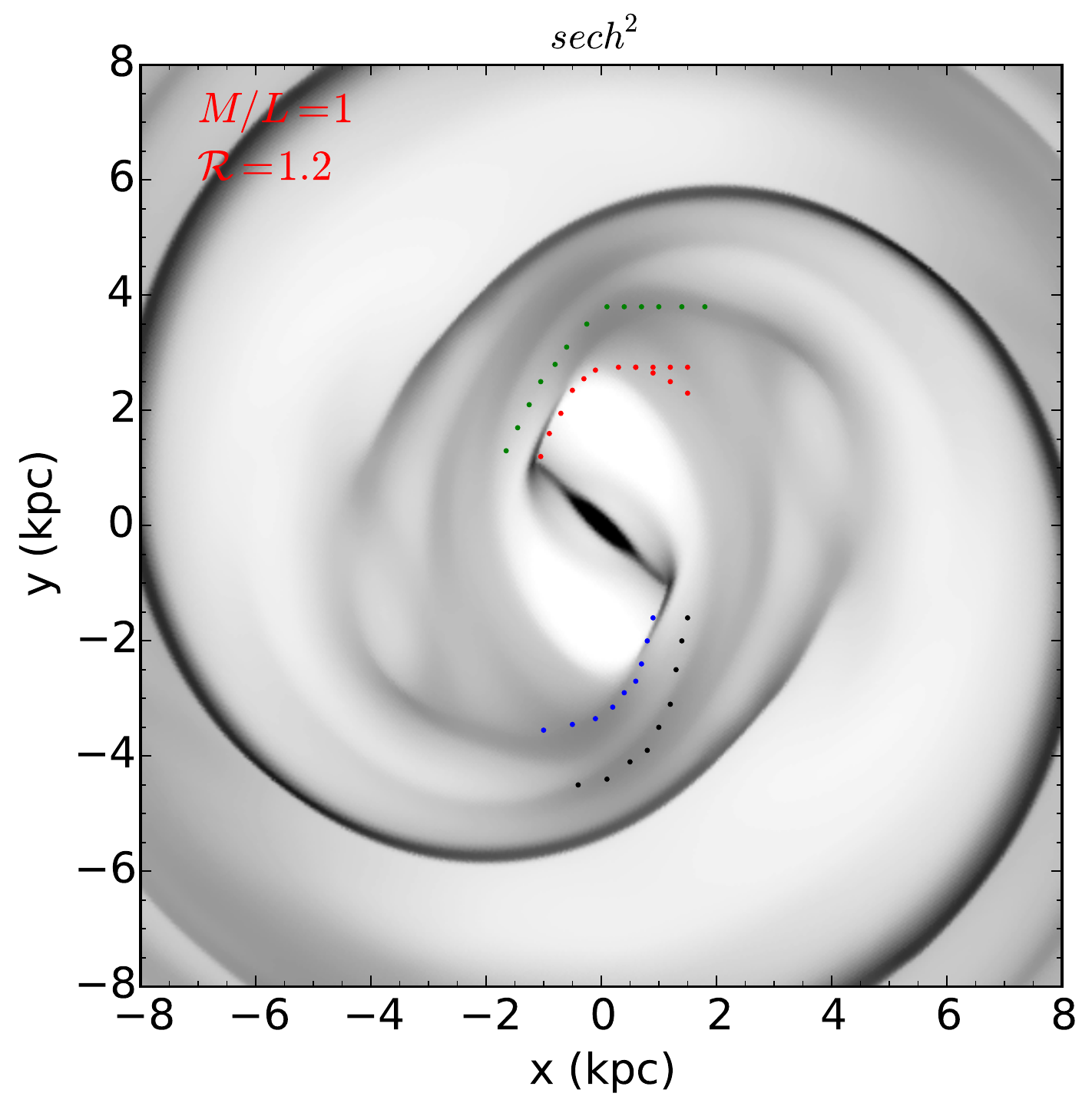}
	\label{fig:sech2dens}}
\quad
\subfigure[Exponential]{%
	\includegraphics[width=0.22\textwidth]{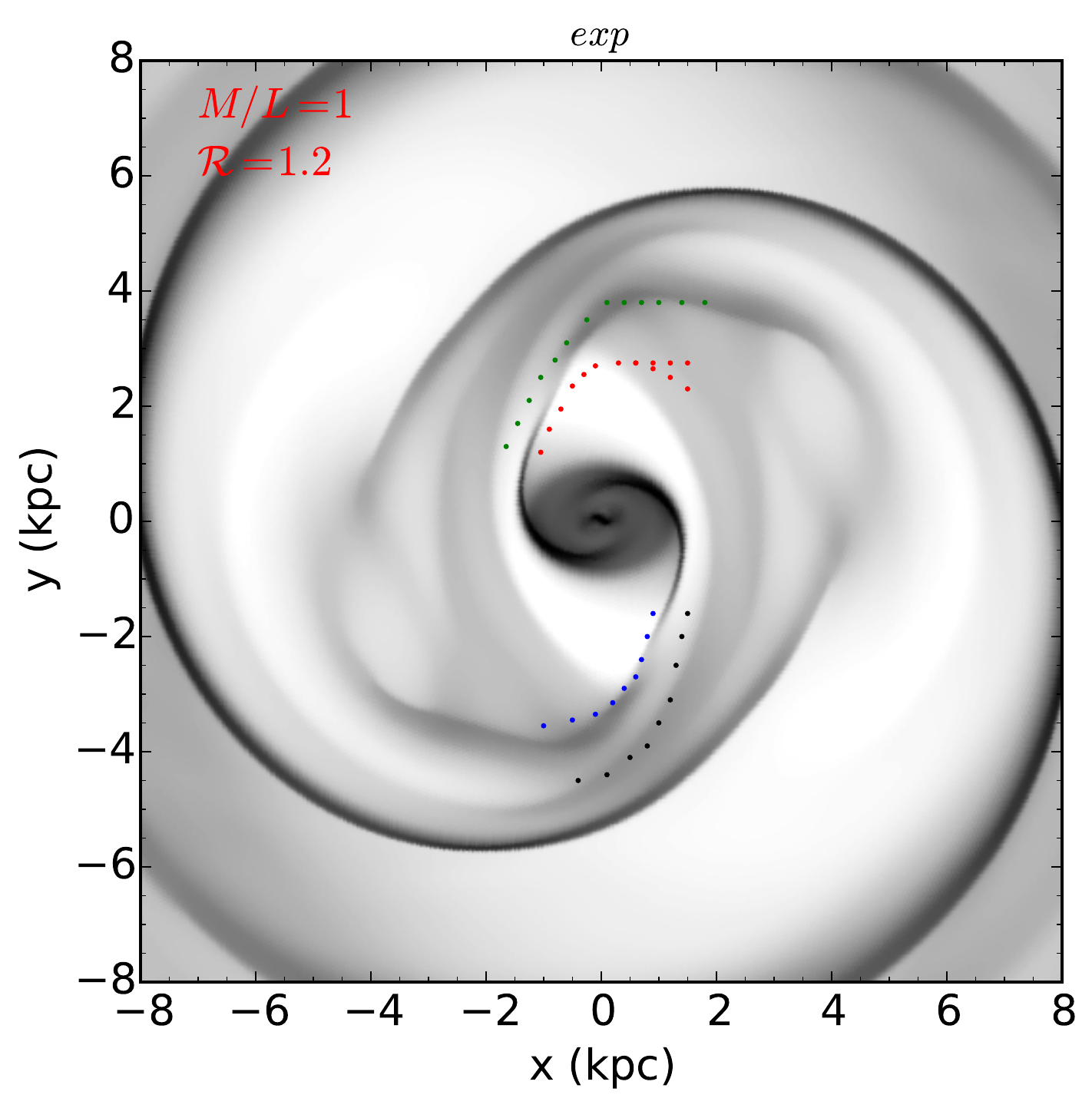}
	\label{fig:expdens}}
\quad
\caption{Comparing the gas morphology of models with an isothermal (left) and exponential (right) height function.} 
\label{fig:expsech}
\end{figure}

\begin{figure*}
\centering
\includegraphics[width=0.58\textwidth]{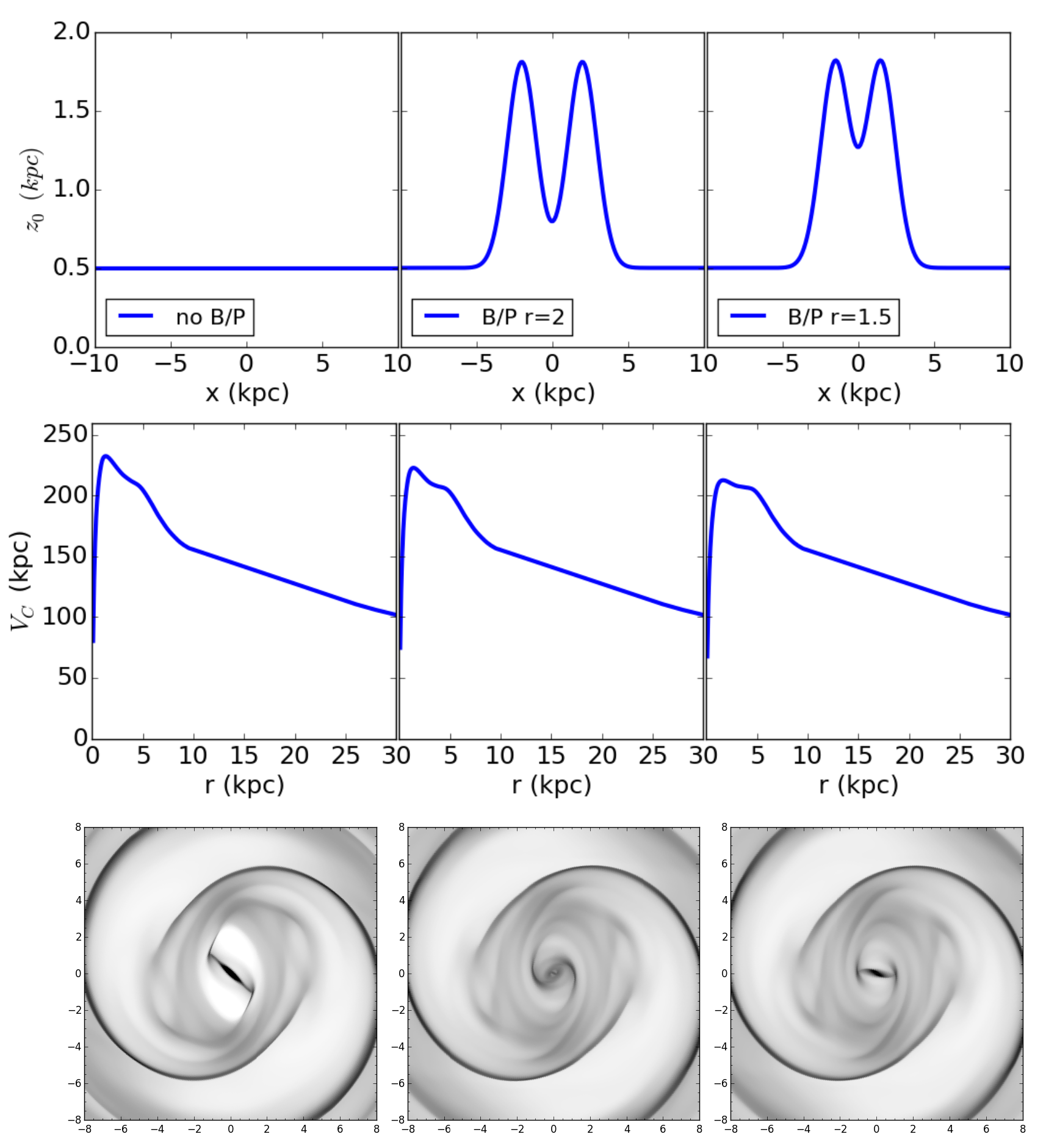} 
\caption{Effect of changing the length of the b/p bulge on the disc rotation curves and gas response of the models: \emph{Top row:} The variation of the scaleheight of the model along the $x$-axis for model 001 which does not contain a b/p bulge, model 034 which contains a b/p bulge with length 2\,kpc and model 066 which contains a b/p bulge with length 1.5\,kpc (left, middle and right columns respectively). \emph{Second row:} The rotation curve of the three models. \emph{Bottom row:} The gas surface density of the models.}
\label{fig:vc_bp}
\end{figure*}

By increasing the scaleheight of the disc, we reduce the density in the plane of the galaxy, which effectively reduces the strength of the forces in the plane. This results in weaker shocks in the gas, since the ratio of the non-axisymmetric-to-total forcings is also smaller. This can be seen in Figure \ref{fig:heightfunction} where we plot the gas response simulations in five models with increasing scaleheight (where the scaleheight of the model is noted in the upper left corner of the plot). The shocks become rounder, less extended, and there is more gas present in the bar region, since the bar cannot deplete the gas. There is also less of a central concentration in the responding gas in the models with thicker discs, since, due to the weakening of the bar, less gas is funnelled to the central regions.

We also explore the effects of an exponential height function on the shocks, since it is possible that by assigning an exponential height function to the models, the results could be altered. We show in Figure \ref{fig:expsech} the effects of assuming an exponential height function for the models as compared to an isothermal height function, where both models have an equivalent scaleheight (the exponential scaleheight $h_z$ = $z_0$/2 where $z_0$ is the isothermal scaleheight.) As expected, the shocks in the exponential model are slightly stronger. This occurs because for the exponential height function the density distribution is more peaked in the mid-plane of the galaxy (see Figure \ref{fig:scaleheights}), which results in larger forces in the plane of the galaxy, and hence stronger shocks. After running a large number of models with an exponential height function we found that the range of acceptable values for M/L, $r_L$ and $h_z$ is similar for both the exponential and isothermal height functions (see Section \ref{sec:bestfit}).

When a b/p height function is added to the model, the reduction of the non-axisymmetric forces is localised around the area where the b/p is maximum; this occurs due to the thickening of the disc mainly at the locations where the b/p is maximum. We model two different b/p geometries, where the maxima occur at $r$=2 and 1.5\,kpc respectively along the bar. In Figure \ref{fig:vc_bp} we show the effect of adding these b/p bulge models on the rotation curve of the disc (second row), where we see that the b/p bulges reduce the height of the rotation curve in the region where the b/p is maximum. 
The effect of the b/p bulge on the shape of the shocks is also evident in the gas response to these models, as can be seen in the bottom row of Figure \ref{fig:vc_bp}, where the shocks become rounder and have discontinuities. Also the area within the bar region is not as depleted of gas as the model without the b/p, and there is less gas in the central regions of the model with the b/p bulge (for more details on the gas inflow in models with and without b/p bulges see \cite{Fragkoudietal2016}.)

From Figures \ref{fig:heightfunction} and \ref{fig:masstolight} we can see how increasing the scaleheight of the disc has a similar effect on the gas flow as decreasing the M/L; the shocks become weaker and rounder. Therefore we see that there is a threefold degeneracy i.e. there is a degeneracy between the height function, the M/L, and the Lagrangian radius of the models, although the degeneracy between the M/L and scaleheight is strongest. This is one of the difficulties of this type of study, and one of the ways we deal with these degeneracies is by using the velocity dispersion method in order to restrict the allowed parameter space, as will be discussed in the next subsection. 

\subsection{Best fit models}
\label{sec:bestfit}
\begin{figure*}
\centering
\subfigure[]{%
	\includegraphics[width=0.43\textwidth]{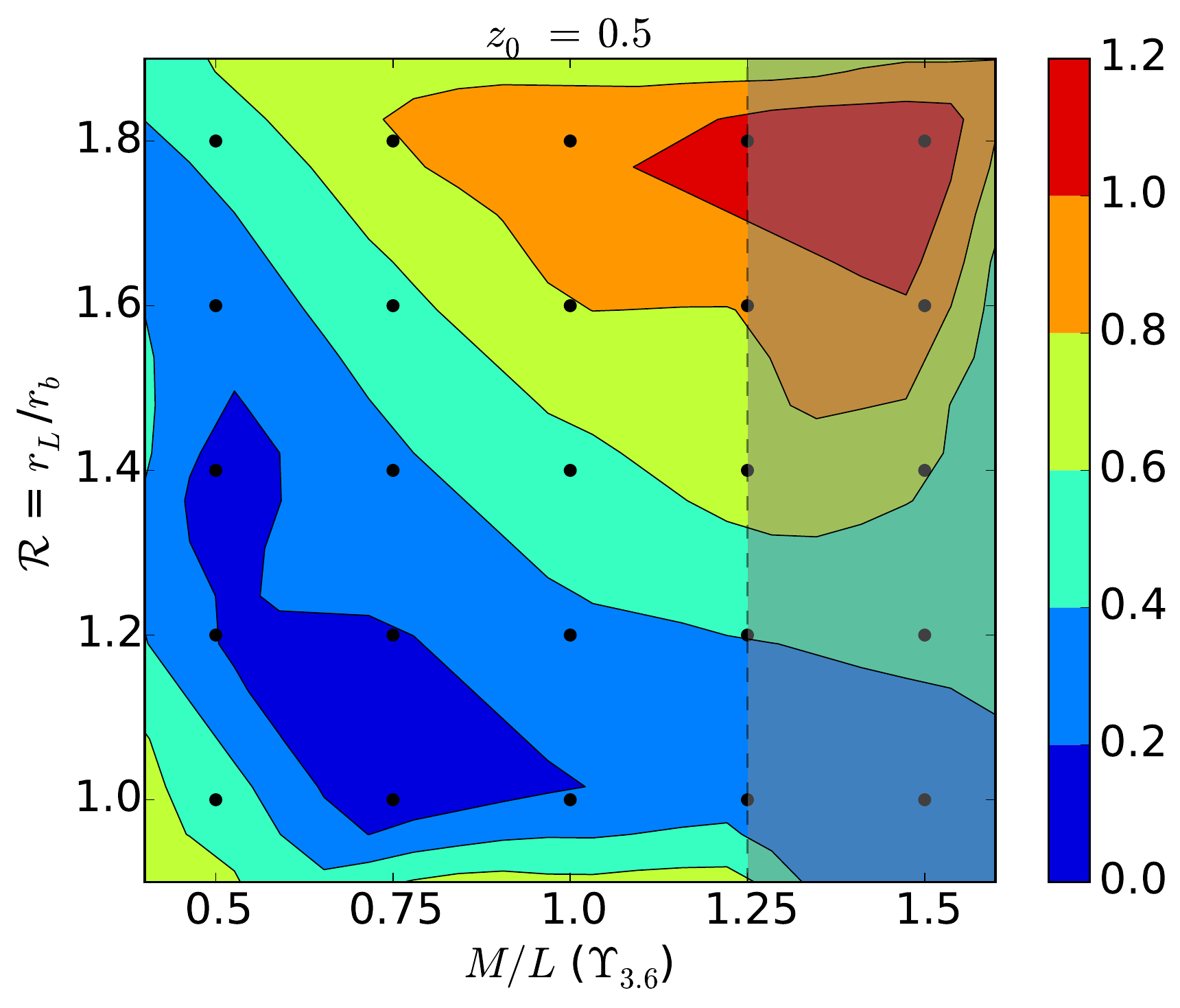}
	\label{fig:cont05}}
\quad
\subfigure[]{%
	\includegraphics[width=0.43\textwidth]{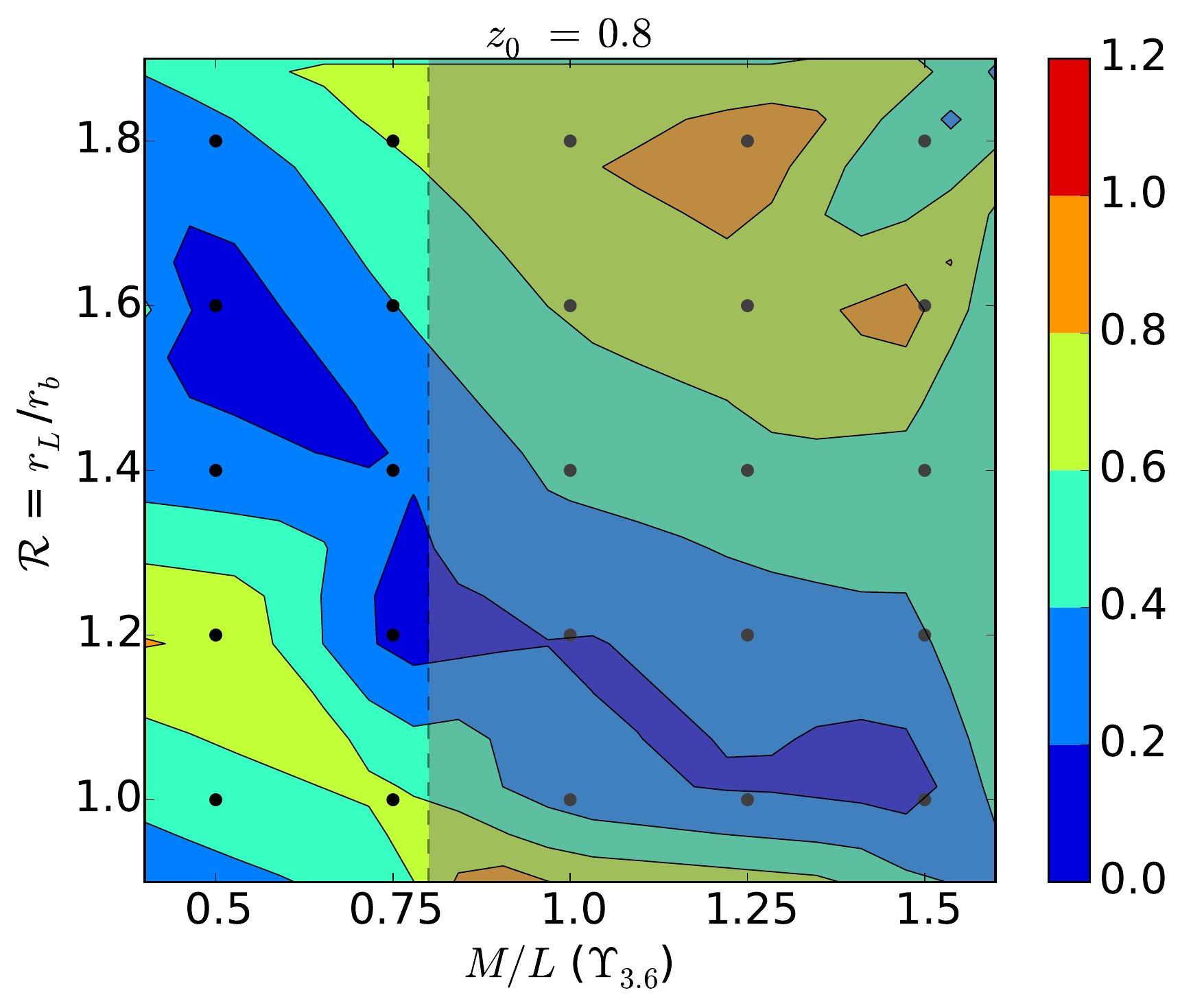}
	\label{fig:cont08}}
\quad
\subfigure[]{%
	\includegraphics[width=0.43\textwidth]{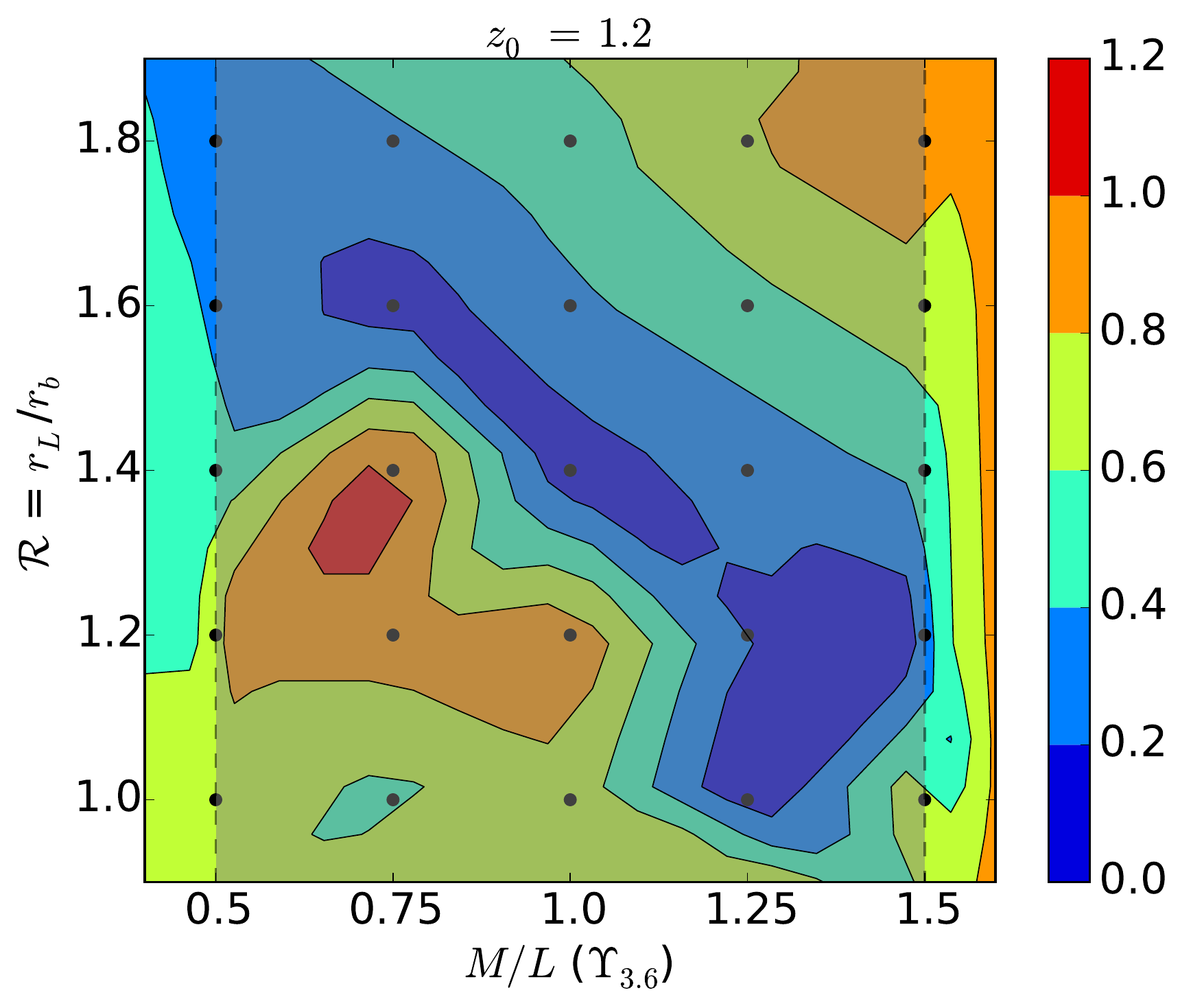}
	\label{fig:cont12}}
\quad
\subfigure[]{%
	\includegraphics[width=0.43\textwidth]{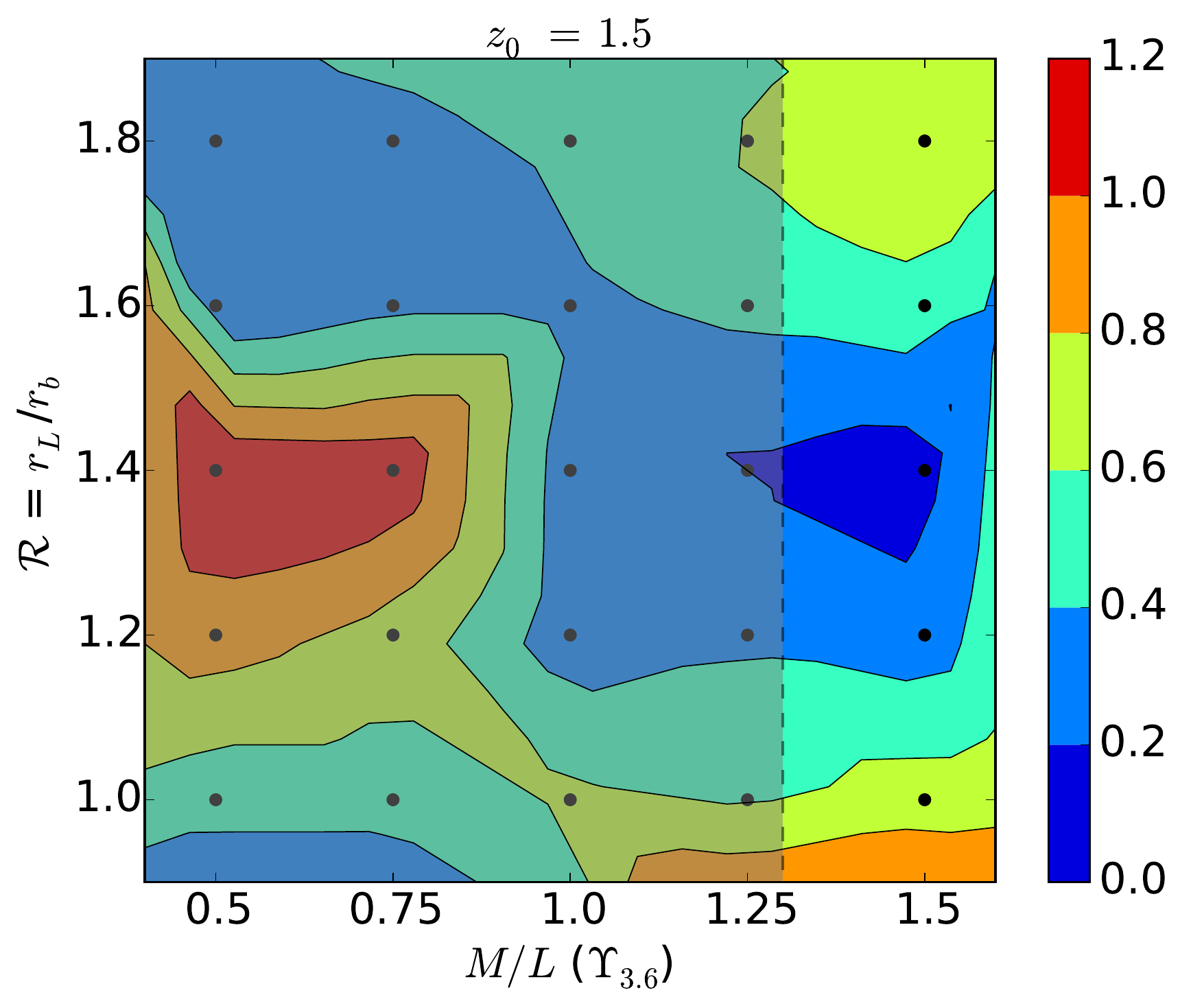}
	\label{fig:cont15}}
\quad
\subfigure[]{%
	\includegraphics[width=0.43\textwidth]{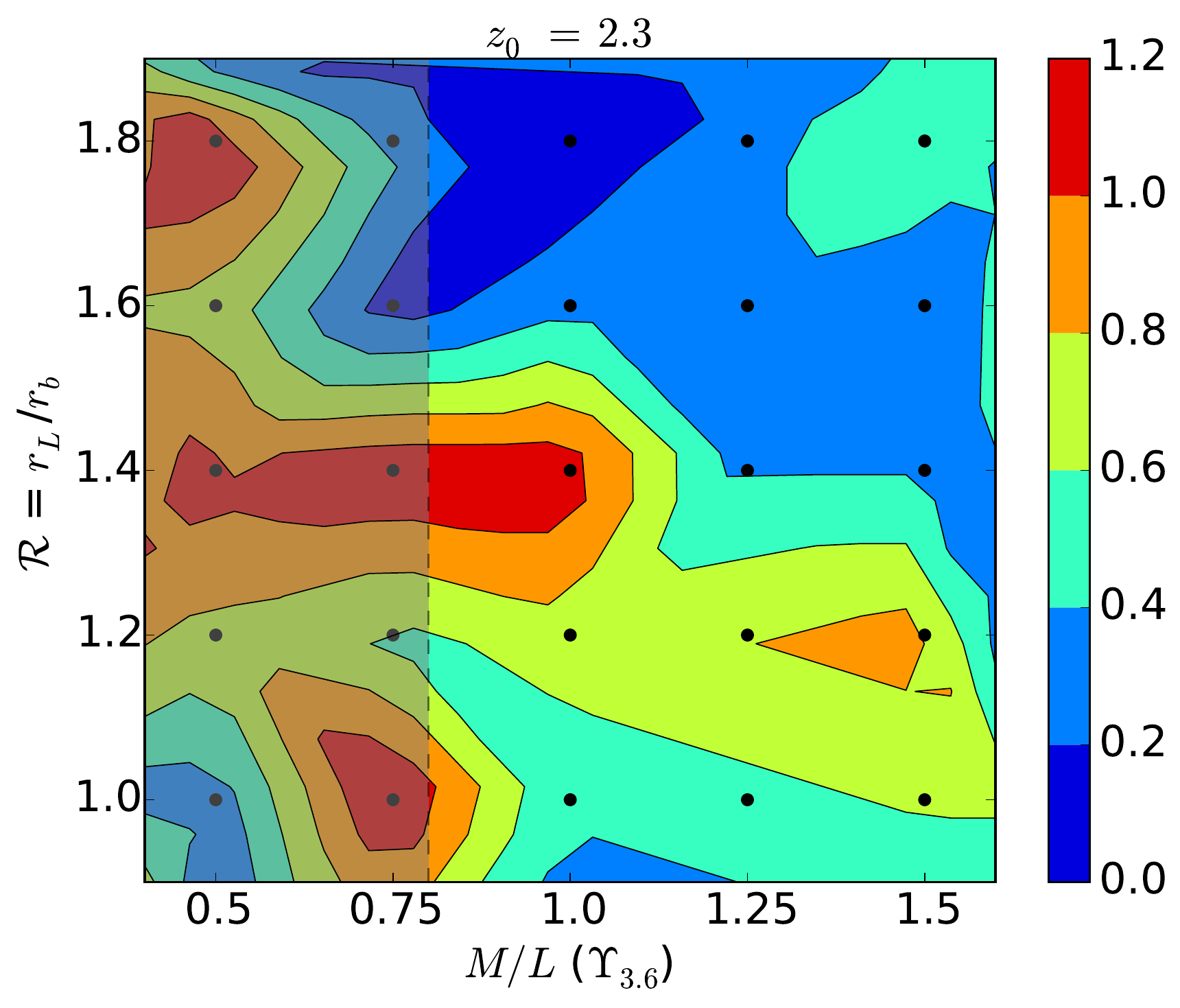}
	\label{fig:cont23}}
\quad
\subfigure[]{%
	\includegraphics[width=0.43\textwidth]{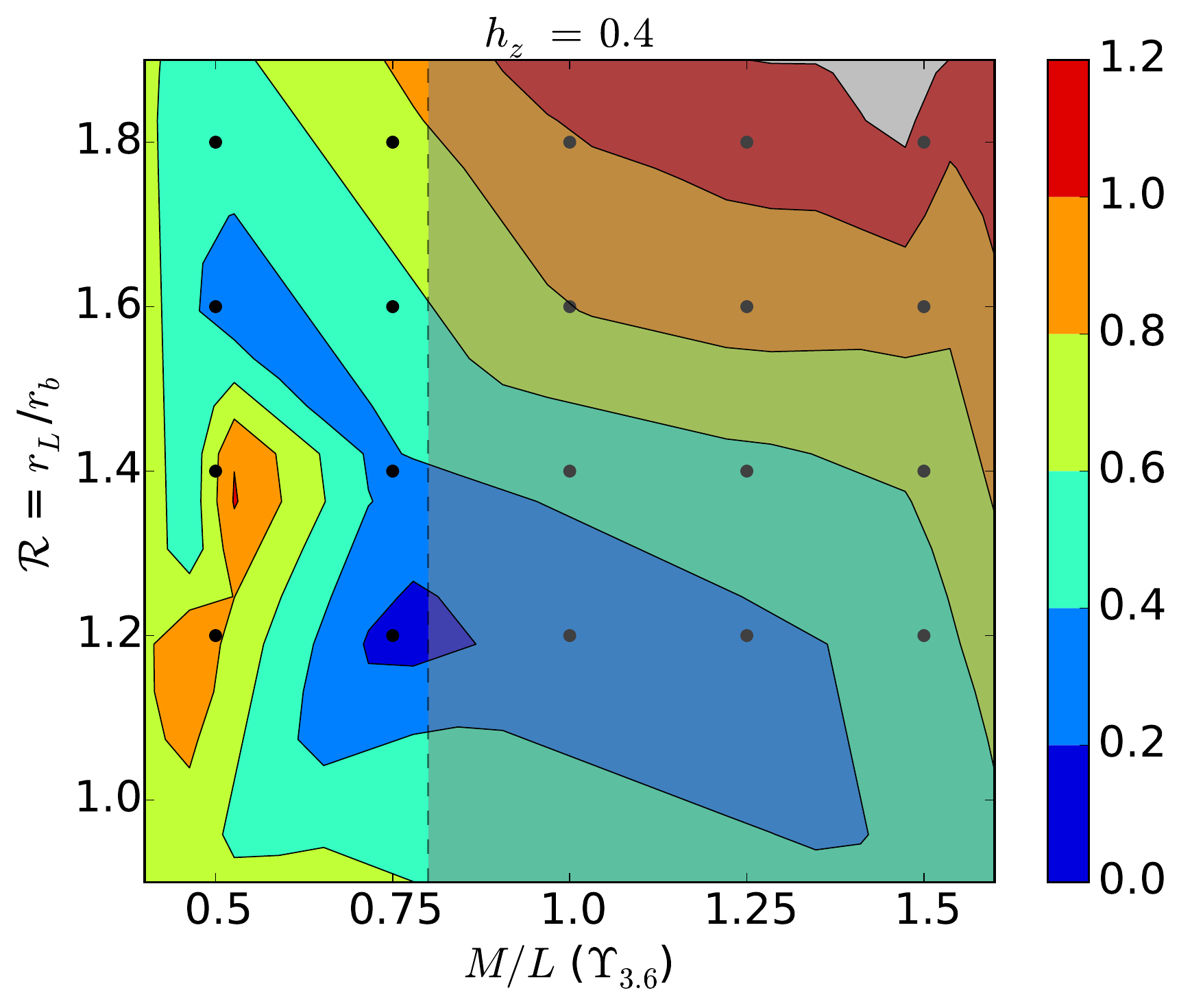}
	\label{fig:conthz05}}
\quad
\caption{Contour plots of $\Delta l$, (i.e. the goodness of fit) for suites of models with different scaleheights (the scaleheight is denoted on the top of each plot).The bottom right contour plot corresponds to an exponential height function with $h_z$=0.4\,kpc (for the sake of brevity, we do not include all the suites of models run with an exponential height function). The $x$-axis of the plots correspond to M/L (in units of $\Upsilon_{3.6}$). The $y$-axis gives $\mathcal{R}$, i.e. the ratio of the Lagrangian radius to bar semi-major axis, $r_L$/$r_B$. The black dots correspond to points on the grids where we have models. The shaded area of the plots shows the constrained parameter space after taking into account the vertical velocity dispersion of the galaxy (see text).} 
\label{fig:allbiggrids}
\end{figure*}

\begin{figure*}
\centering
\includegraphics[width=0.8\textwidth]{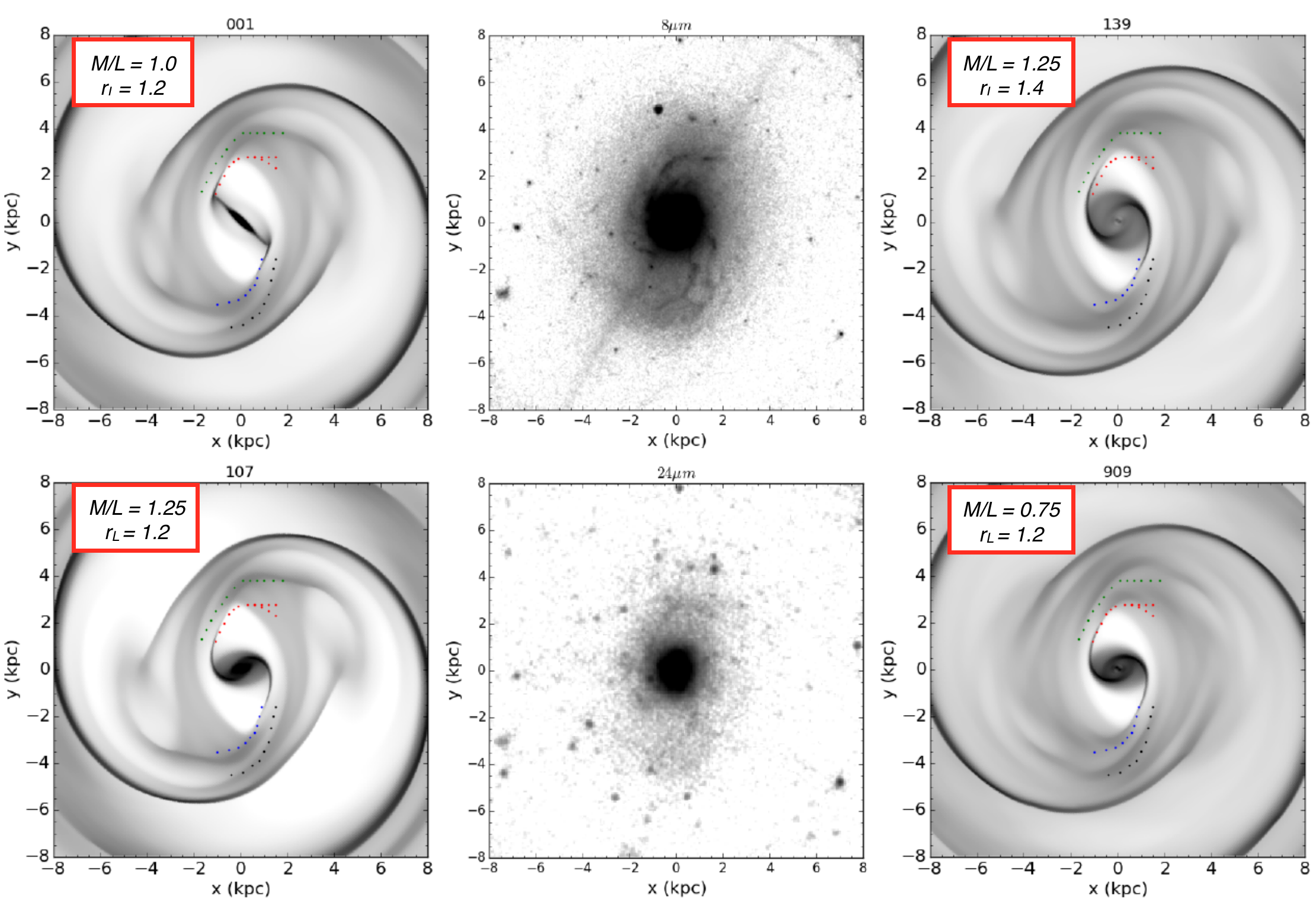}
\caption{Two dimensional gas surface density for the best fit models. In the left column the top row shows model 001 and the bottom row 107 which have $z_0$=0.8\,kpc, M/L = 1.0 and 1.25$\Upsilon_{3.6}$ respectively and $r_L$=1.2. The second column shows the 8 and 24$\mu$m images of NGC~1291 for comparison. In the left column, the top row shows model 139 which has $z_0$=0.8\,kpc with a b/p with length=1.5\,kpc along the bar, M/L=1.25 and $r_L$=1.4, while the bottom row shows model 969 which has an exponential height function with $h_z$=0.4\,kpc, M/L=0.75 and $r_L$=1.2 ($r_L$ is given in units of $r_B$).} 
\label{fig:bestfit}
\end{figure*}

\begin{table}
\centering
\begin{tabular}{ c | c | c | c | c | c | c } 
Model & M/L ($\Upsilon_{3.6}$) & $\mathcal{R}$ & HF & $z_0$ (kpc) & Max & b/p \\ \hline \hline
001 & 1 & 1.2 & iso & 0.8 & 74\% & no \\
107 & 1.25 & 1.2 & iso & 0.8 & 80\% & no \\
139 & 1.25  & 1.4 & iso & 0.8 & 80\% & yes \\
909 & 0.75  & 1.2 & exp & 0.8 & 63\% & no\\
\end{tabular}
\caption{Best fit models: From left to right: the model number, the M/L (in units of $\Upsilon_{3.6}$), the ratio of the Lagrangian to bar radius, the type of height function used (either isothermal or exponential), the scaleheight $z_0$ in kpc (or the equivalent $h_z$ in the case of an exponential profile), the contribution of the disc circular velocity to the total circular velocity at $r$ = 2.2$r_h$ (above 75\% is maximal), and whether or not the model contains a b/p bulge. }
\label{tab:bf}
\end{table}

Out of the different suites of models that were run we construct a number of grids for which we create contour plots showing the value of the distance $\Delta l$, between the dust lane in the 8$\mu$m and 24$\mu$m images and the maximum of the density in the gas dynamic simulations. For a given scaleheight and height function, we make a grid of models with different M/L and different Lagrangian radii (i.e. pattern speeds) as is shown in Figure \ref{fig:allbiggrids}. 

To find the best fit models to the observations we followed a procedure of elimination, by applying the following criteria: 
\begin{enumerate}
\item The location of the primary shocks should be as close as possible to the dust lanes in the image (determined as shown in Figure \ref{fig:mod005_slit}). 
\item The model should have a \emph{dominant} strong shock, i.e. one which is stronger than any secondary shocks formed in the gas flow.
\item The shape of the shocks should match the dust lanes in the 8$\mu$m image and should also reproduce features in the 24$\mu$m image.
\end{enumerate}

We choose the models with the smallest $\Delta l$, from the contour plots in Figure \ref{fig:allbiggrids}, from the allowed parameter space of models. This reduces the number of best fit models to 20 models. We then examine the models visually in order to determine the goodness of fit according to the set criteria.

Additionally, we combine this method with the velocity dispersion method, in order to help break any remaining degeracies. We use the relation between the scaleheight, velocity dispersion and surface density of an isothermal sheet in order to constrain the relation between the scaleheight and the M/L. This has been used in a number of studies in the past (e.g \citealt{Bottema1993, Bershadyetal2010,Martinssonetal2013}). Assuming the disc is isothermal, the relation is given by: 
\begin{equation}
z_0 = \frac{\sigma_z^2}{\pi G \Upsilon I}
\label{eq:sigma}
\end{equation}
where $I$ is the surface intensity, in units of flux per unit area and $\Upsilon$ is the M/L \citep{BT2008,Bershadyetal2010}. 

From \cite{Bosmaetal2010} we take a value for the stellar velocity dispersion corresponding to the location of the pseudo-slit. This allows to constrain the parameter space, and  exclude some of the models which have a M/L incompatible with the scaleheight predicted by the relation.
It is possible that in the bar region the disc is not exactly isothermal, which is why we allow for a range of values ($\pm$25\%) for the M/L given a certain scaleheight. Thus this relation can exclude some extreme cases in which for an adopted M/L the scaleheight is very far off from that given by the theoretical value of the above relation. This allowed region is given by the gray shaded region in the plots in Figure \ref{fig:allbiggrids}.

The four best fit models are shown in Figure \ref{fig:bestfit} together with the images of NGC~1291 in 8 and 24$\mu$m for comparison. They all have a scaleheight of 0.8\,kpc (equivalent to 0.4\,kpc for the models with an exponential height function), while models which are even thinner than this tend to give a worse fit. In general we expect the thickness of the disc to be between 0.8-1.2\,kpc since also for 1.2\,kpc the models give worse fits than for 0.8\,kpc. In all the models, fast bars are preferred with a value of $\mathcal{R}$ $\leq$ 1.4 since for slower bars the shocks tend to generally be too offset from the bar to correspond to the locations of the dust lanes. 

The best fit models are summarised in Table \ref{tab:bf}. They are: 001 and 107 which correspond to $z_0$=0.8\,kpc without a b/p bulge and have M/L=1 and 1.25$\Upsilon_{3.6}$ respectively and Lagrangian radius $r_L$=1.2. Model 139 is also a relatively good fit to the dust lanes and has a b/p bulge with a length of 1.5\,kpc along the bar, with a M/L=1.25$\Upsilon_{3.6}$ and $r_L$=1.4. Additionally, model 909 with an exponential height function and $h_z$=0.4\,kpc (equivalent to $z_0$=0.8\,kpc), M/L=0.75 and $r_L$=1.2 also gives a good fit to the observations.

It should be noted, that none of the models give a \emph{perfect} fit to the morphology of the dust lanes.
This could be due to a number of reasons, such as the fact that we symmetrise the model although the galaxy is not perfectly symmetric, or due to small effects from inclination (i.e. even though we assume the galaxy is face-on, we know that it has a small, but perhaps non-negligible inclination). Additionally, although a large number of simulations were run, the grid we explore is still quite ``coarse'' and there are regions of the parameter space which have not been explored in detail. It is possible also, for example, that there is a boxy, rather than a strongly peanut shaped, bulge.

\section{Discussion}
\label{sec:discussion4}

\subsection{The height function of NGC~1291}
The best fitting models predict a disc with scaleheight $\approx$0.8\,kpc for NGC~1291.
The thicker the scaleheight of the model, the larger the M/L that needs to be adopted, in order to be able to reproduce the shocks in the gas. However, due to the fact that there is an inverse proportional relation between the scaleheight and the M/L via the vertical velocity dispersion, the M/L cannot be arbitrarily increased for a thick disc, since this would lead to a higher stellar velocity dispersion than that which is observed. 
Additionally, by increasing the scaleheight of the disc, the shape of the rotation curve of the model changes; therefore, even by increasing the M/L, the shock loci of the gas do not reproduce well the the shapes of the dust lanes, which become too round for thicker discs.

An interesting point to note is that for borderline maximal disc cases such as NGC 1291, the functional form of the height function (i.e. whether we use an exponential or an isothermal) can have a significant effect on the results. Using an exponential can lead to a lower M/L, while using an isothermal can lead to a higher one. This, combined with the fact that the scaleheight can affect the results even more than the functional form, shows that a lot of attention must be paid to the vertical structure of the disc when carrying out this kind of study.

NGC~1291 seems to have a barlens, which hints at the existence of a b/p bulge \citep{Athanassoulaetal2014,Laurikainenetal2014}. However, out of the best fit models shown in Figure \ref{fig:bestfit}, only one contains a b/p bulge.
There is some evidence that the shocks in the \emph{central-most} part of the galaxy are rounder than the shape of the gas shocks for models without a b/p bulge. This hints at the possibility that there is a thicker structure in the central region, one which weakens the shocks thus making them rounder.\footnote{As shown in the Appendix one should be cautious when examining the central region of the simulations, since this region is dependent on the grid resolution of the simulations.} It is therefore possible that a thicker region, for example due to a boxy shaped bulge, rather than a strong peanut or X-shaped bulge, exists in NGC~1291.

Indeed, since we have only explored a small fraction of the parameter space for parameters of the b/p bulge, such as the length, width and height of the b/p, it definitely cannot be excluded that NGC~1291 could contain a weak b/p bulge. However the results seem to suggest that the galaxy does not contain a \emph{strong} b/p bulge, i.e. one with a large double peaked maximum, however a weaker peanut or boxy shaped bulge, or indeed a peanut bulge with smaller length and width, cannot be excluded.

\subsection{Dynamical determination of the 3.6$\mu$m M/L}
The best fit models obtained in this study have a M/L which falls within the range predicted by SPS models and other approaches, which predict a value of $\Upsilon_{3.6}$=0.6M$_\odot$/L$_\odot$ (assuming a Chabrier IMF) with small variations for age and metallicity \citep{Meidtetal2014,Norrisetal2014,Roecketal2015}. 

SPS methods are highly dependent on the assumed IMF -- indeed the choice of IMF can change the M/L by up to a factor 10 \citep{BelldeJong2001} and in more recent works by up to a factor of 4 \citep{Roecketal2015}. On the other hand, since our technique provides a dynamical determination of the M/L, and as such is independent of IMF, it provides an independent confirmation of the robustness of these methods. Our best fit models have a M/L which lies in the range of M/L$_{3.6}$=0.6-0.75, in line with the uncertainty (i.e. about 0.1dex) present in the above mentioned methods \citep{Meidtetal2014}.

\subsection{Maximal or sub-maximal disc?}

\begin{figure}
\centering
\includegraphics[width=0.45\textwidth]{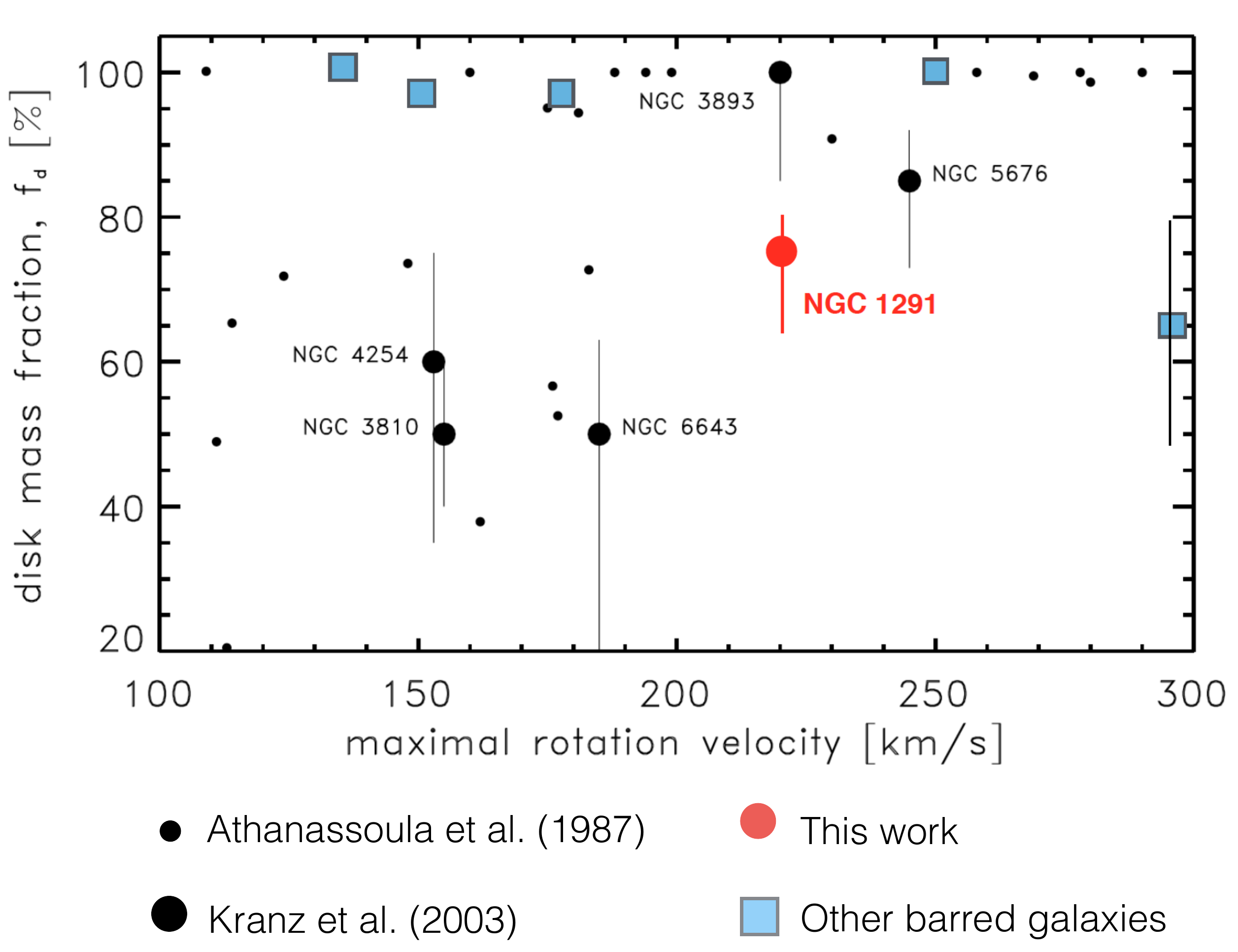}
\caption{Plot of the disc mass fraction as a function of maximum rotation velocity. We translate our value of $f_d$ to the equivalent of that used in Kranz et al. (2003), and add it to the plot (red circle). We see that NGC~1291 falls on the upper right side of the plot, consistent with galaxies with $V_{max}\geq 200$ having maximal discs. The blue squares correspond to previous works on the M/L of barred galaxies using gas response simulations, i.e. Lindblad et al. (1996), Weiner et al. (2001), Weiner et al. (2004) and Zanmar-Sanchez et al. (2008). Adapted from Kranz et al. (2003).} 
\label{fig:kranzcomp}
\end{figure}

We examine the rotation curves of all the best fit models, to explore whether the disc of NGC~1291 is maximal. Model 001 has a disc which contributes 74\% of the rotation curve of the galaxy at $r$ = 2.2 $h_r$. Therefore according to the definition by \cite{Sackett1997} the disc of this galaxy is just short of maximum. Both models 107 and 139 have a disc which contributes 80\% of the total circular velocity at $r$=2.2$h_r$, which therefore makes them both maximal. The disc of model 909 contributes 63\% of the total rotation velocity at $r$=2.2$h_r$, and thus would be dubbed sub maximal. However, it is worth noting that the scalelength of NGC~1291 is unusually large (5.8\,kpc); thus even for model 909, the disc contributes most of the circular velocity in the inner regions of the galaxy and at least until the end of the bar (which is at $r$=5.1\,kpc). Thus, all our best fit models seem to be consistent with a maximal disc in the inner regions of the galaxy, and at least up to the end of the bar, although model 909 is slightly sub-maximal according to the definition by \cite{Sackett1997}.

These results are consistent with most previous results in the literature using a similar approach, i.e.  gas dynamical modelling. In the work by \cite{Weiner2001}, they show that the galaxy they study, NGC 4123 ($V_{max}$$\approx$140\,km/s), has a maximum disc. \cite{Zanmar-Sanchezetal2008} find that the disc of NGC 1365 ($V_{max}$$\approx$300\,km/s) is massive although not quite maximal. The work of \cite{Kranzetal2003} showed that the two most massive galaxies in their sample (those with $V_{max}$$\geq$200\,km/s) had maximal discs, while the less massive galaxies were sub maximal (see Figure \ref{fig:kranzcomp}). It seems therefore that there is some tension between the results found in \cite{Weiner2001} and those in \cite{Kranzetal2003}, since NGC 4123 should have a submaximal disc according to the findings of \cite{Kranzetal2003}. We are consistent with these findings, i.e. the disc of NGC~1291, which is a galaxy with $V_{max}$$\geq$200\,km/s, is maximal. This seems to argue that massive barred galaxies tend to have massive discs as most of the aforementioned studies also find (e.g. \citealt{ABP1987}) . Whether or not this also extends down to lower masses remains an open question. 

Our results are inconsistent with some other methods for dynamically determining the M/L, such as for example the DiskMass survey \citep{Martinssonetal2013}, which make use of the stellar vertical velocity dispersion. In their study they found that all of the 30 galaxies used for the study have submaximal discs, even though about a third of their sample are spiral galaxies with $V_{max}$$\geq$200\,km/s. This incosistency could be due to the fact that the velocity dispersion probed in these studies corresponds to a younger (and therefore colder) stellar population, which leads to a smaller velocity dispersion and subsequently a smaller M/L \citep{Aniyanetal2016}.
On the other hand, we are consistent with the findings of \cite{Bosmaetal2010} for NGC~1291 who found, by exploring the vertical velocity dispersion as well as the photometric properties of the bar, that the disc of the galaxy is maximal.

\subsection{Fast or slow rotating bar?}
\label{sec:fastslowbar}
The three best fit models for NGC~1291 have fast rotating bars with $\mathcal{R}$ $\leq$ 1.4. This is consistent with observational estimates of the bar pattern speed for early type galaxies \citep{Elmegreen1996,Corsini2011}, although there are indications that bars in late type galaxies might rotate slowly \citep{Rautiainenetal2008}. The results are also in agreement with theoretical values predicted by \cite{Athanassoula1992b}, who used the shape of the dust lanes to constrain $\mathcal{R}$ and found that it should be of the order of 1.2$\pm$0.2.

Additionally, this is consistent with previous gas-dynamical studies of other barred galaxies, which are also found to have fast rotating bars, such as NGC 4123 \citep{Weiner2001} and NGC 1365, which was found to have a fast rotating bar by two independent studies  ($r_L$$\approx$1.2$r_B$, \cite{LLA1996,Zanmar-Sanchezetal2008}).

This result is also consistent with the fact that NGC~1291 is compatible with the maximum disc hypothesis, since there are theoretical arguments which suggest that galaxies embedded in concentrated halos will lose angular momentum and thus slow down \citep{Athanassoula2003}.

\subsection{Additional physical processes}
\label{sec:physics}
The simulations used in this study are kept as ``stripped-down'' as possible in order to avoid significant computational costs, due to the large number of simulations which need to be run in order to explore the three dimensional parameter space as finely as possible. We therefore trade detailed physics (such as gas cooling, self-gravity, star formation, feedback etc.), three dimensions and very high resolution in order to be able to run a larger number of simulations. Here we briefly discuss the effects of leaving out these physical processes on our results.

As previously mentioned, and as is shown in our resolution study in Appendix A, the resolution should not play a major role on the shape and loci of the shocks (to within 5\% accuracy). However, as was already shown by \cite{Sormanietal2015}, the central regions are affected quite severely by the grid resolution, which is why we exclude these regions from the analysis in this work. 

\cite{Fujimotoetal2014} show that in three dimensions, cold gas in the bar region can have significant vertical velocity dispersions (see their Figure 4). This vertical displacement could indeed have an effect on the shape of the shocks; however, we expect this effect to be minimal, since the radial and tangential velocities in the disc are higher than the vertical velocity dispersion in the gas (see for example Figures 10 and 11 in \cite{Athanassoula1992b}). 

By including more detailed physics in our simulations, such as gas cooling and self-gravity, the gas would fragment and develop clumps \citep{Nimorietal2013,Fujimotoetal2014}, while star formation and feedback from supernovae can then disperse these dense molecular clouds \citep{Dobbsetal2012,Renaudetal2013}. While including these recipes into our simulations would change the detailed structure of the shocks in the dust lane, we don't expect the shock loci to be affected globally. Additionally the correct implementation of star formation and feedback recipes is a complex issue, well beyond the scope of this paper.

Therefore, considering the large amount of simulations needed to explore the parameter space, we think that the reduced realism of our simulations will not compromise the conclusions of this work; however, we plan to further explore the effects of more detailed physics and three dimensional gas flows in future work.

\section{Summary}
\label{sec:summary4}

We explore the dynamical structure of NGC~1291, by using the gas in the galaxy as a tracer of the underlying gravitational potential, in order to constrain the disc M/L and therefore the amount of dark matter in its central regions. The potential of the galaxy is obtained from the 3.6 and 4.5$\mu m$ Spitzer images, and has three free parameters, namely the height function of the galaxy, the M/L and the bar pattern speed (or equivalently, the Lagrangian radius). We run hydrodynamic gas response simulations in these potentials, and explore the three dimensional parameter space, to obtain the best fit models, by matching the shock loci in the gas response simulations to the loci of the observed dust lanes in the 8 and 24$\mu$m images. 

There are a number of interesting trends which emerge from this study:
the effects of the scaleheight and functional form of the height function on the gas flows in hydrodynamic gas response simulations have not been extensively studied in the past. We show that by increasing the scaleheight of the disc, or by employing a less peaked height function, the forces in the plane of the galaxy are reduced, which leads to weaker shocks with a rounder morphology.  This effect is similar to that which occurs when the M/L of the model is decreased, while increasing the contribution of the dark matter halo to maintain the outer part of the rotation curve at a value predicted by the TF relation. In both cases, i.e. when we increase the scaleheight or decrease the M/L, we are decreasing the strength of the non-axisymmetric forcings in the plane of the galaxy -- albeit in a slightly different way in each case. Furthermore, in accordance with previous results in the literature \citep{Athanassoula1992b,Sanchez-Menguianoetal2015}, we show that by decreasing the pattern speed of the bar, the shocks in the gas are displaced from the bar semi-major axis. Conversely, by increasing the pattern speed of the bar the morphology of the shocks becomes rounder and, for very fast bars, the shocks are weaker.

We find that there are degeneracies between the three main free parameters which make it difficult to unambiguously find the best fit model for the galaxy. We therefore combine the gas response method with the velocity dispersion method, in order to further constrain the parameter space.

The best fit models obtained for NGC~1291 suggest that:
\begin{itemize}
\item The M/L of the disc at 3.6$\mu m$ is within the range given by SPS and other methods (once dust contamination is accounted for), i.e. $\Upsilon_{3.6}$$\approx$ 0.6M$_{\odot}/$L$_{\odot}$ \citep{Meidtetal2014,Norrisetal2014,Roecketal2015}. Since our method does not rely on any assumptions about the IMF, it provides a separate independent measurement of $\Upsilon_{3.6}$. 
\item This M/L suggests that NGC~1291 has a maximal disc (three out of four best fit models have maximal discs). This confirms previous findings using similar methods, i.e. that massive barred galaxies have maximal discs \citep{LLA1996,Weiner2001,Kranzetal2003}. 
\item NGC~1291 contains a fast bar, with $\mathcal{R}$$\leq$1.4, also in agreement with previous results in the literature \citep{Elmegreen1996,Rautiainenetal2008,Corsini2011}.   
\item The sceleheight of NGC~1291 is around the range of 0.8-1.2\,kpc (for a $sech^2$ height function). 
\end{itemize}

The results show that a wealth of information can be extracted from the dynamical modelling of barred galaxies via gas response simulations. We plan to expand this work to include other galaxies from the S$^4$G survey, in order to obtain a larger statistical sample and probe a broader range of masses. This should help shed some light on conflicting results in the literature, especially in the low mass regime, where there remains considerable debate as to whether or not discs are maximal.

\section*{Acknowledgements}
FF would like to thank Miguel Querejeta for helpful discussions on obtaining mass maps from NIR images of galaxies. The authors would also like to thank Dimitri Gadotti for providing us with the structural parameters of NGC 1291. We acknowledge financial support to the DAGAL network from the People Programme (Marie Curie Actions) of the European Union's Seventh Framework Programme FP7/2007-2013/ under REA grant agreement number PITN-GA-2011-289313. We also acknowledge financial support from the CNES (Centre National d'Etudes Spatiales - France). This work was granted access to the HPC resources of Aix-Marseille Universit\'{e} financed by the project Equip@Meso (ANR-10-EQPX-29-01) of the program ``Investissements d'Avenir'' supervised by the Agence Nationale pour la Recherche.

\bibliographystyle{mn2e}
\bibliography{References}


\appendix
\section{Tests on simulations: The effect of grid resolution}
\label{sec:appendixB}
\begin{figure*}
\centering
\includegraphics[width=1.\textwidth]{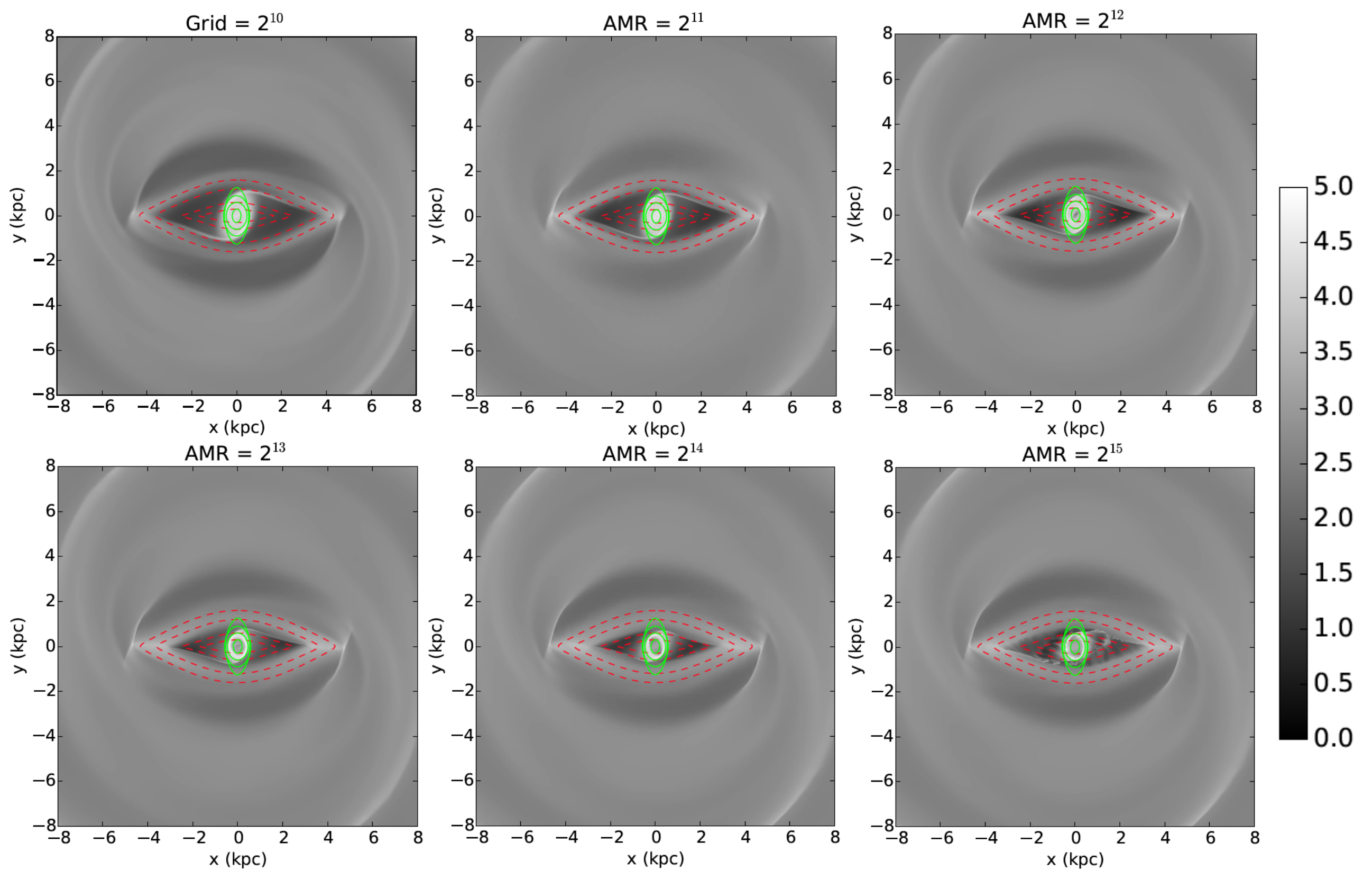}
\caption{Gas surface density with $x_1$ and $x_2$ orbits overplotted. From top left to bottom right we increase the resolution, as noted at the top of each panel. The first panel shows the gas surface density for a cartesian grid with a resolution 40\,pc, where we have a refinement of 2$^{10}$ cells per side. For each subsequent panel we increase the resolution by a factor of 2, and employ an adaptive mesh refinement (AMR) scheme. For the plots titled AMR, the resolution indicated is the maximum resolution reached by the AMR grid.} 
\label{fig:gasorbs}
\end{figure*}

\begin{figure}
\centering
\includegraphics[width=0.49\textwidth]{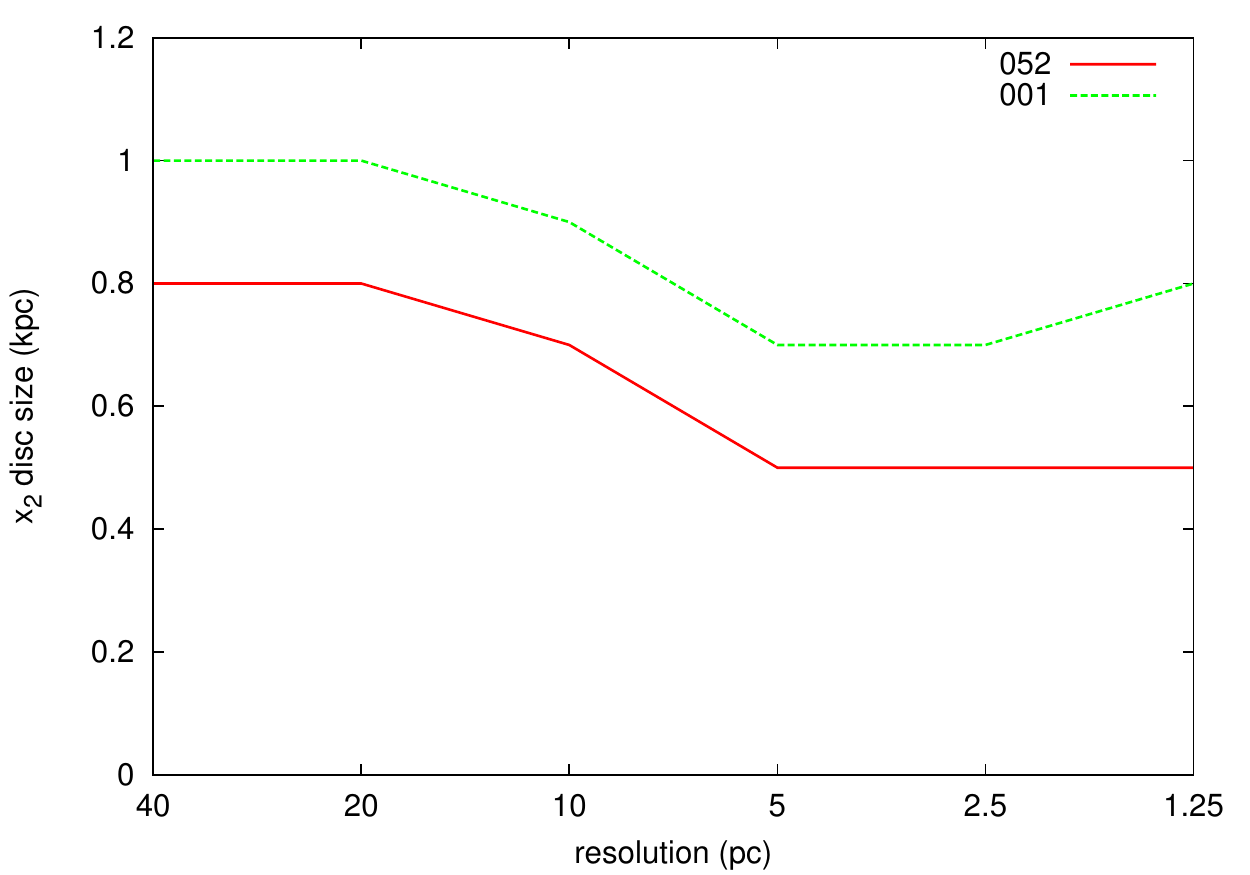} 
\caption{Size of the $x_2$ disc (along the $y$-axis), as a function of grid resolution for two models, $001$ and $052$ from Athanassoula (1992a). Model $052$ has a slightly stronger bar. The size of the $x_2$ disc seems to converge for higher resolutions. For model $001$ the size of the disc seems to increase for higher resolutions but this could be also due to instabilities in the gas.}
\label{fig:discx2converge}
\end{figure}

Recent work by \cite{Sormanietal2015} using gas response simulations, signalled a worrying trend, i.e. that the morphology of the central regions of the gas response simulations is dependent on the adopted grid resolution.
We wanted to explore whether this would affect our ability to draw robust conclusions from the type of analysis carried out in this paper. To do this we ran tests using a number of analytic models containing a disc, bar and bulge, taken from \cite{Athanassoula1992a}. Our aim was to determine the effects of resolution on the shocks induced in the gas flows.

We found, in agreement with \cite{Sormanietal2015}, and as can be seen in Figure  \ref{fig:gasorbs}, that the size of the central disc (sometimes referred to as the ``$x_2$'' disc) is reduced for higher resolutions, presumably due to the gas being able to better resolve the $x_1$ orbits. We measured the extent of the central disc for the different resolutions, and found that the disc size converges for higher resolutions (see Figure \ref{fig:discx2converge}), which is also in agreement with what was found by \cite{Sormanietal2015}. 

We evaluated the effect of resolution on the shock loci, by placing pseudoslits perpendicular to the shocks, similarly to Figure \ref{fig:mod005_slit}. We found that although the central regions of the models are significantly affected, the shock loci are the same for the different resolutions, to within 5\%. This was found for a large number of models.

As can be seen in Figure \ref{fig:gasorbs}, we also found that at high resolution, the gas starts developing instabilities, again in agreement with the findings in \cite{Sormanietal2015}. There is no consensus as to why the flow becomes unstable at high resolutions, but it is a feature common to all hydrodynamic codes, and could be connected with the onset of turbulence due to high shear \citep{Kimetal2012}.

We found therefore that for the purposes of this study, the shock loci are a more robust diagnostic than the central $x_2$ disc, since it is not dependent on the resolution of the simulations. For high enough resolutions, however, the central gaseous disc seems to also converge and therefore, for high enough resolutions, this could also be used as another constraint when comparing gas-dynamical models with observations.

\label{lastpage}

\end{document}